# Pervasive Healthcare: A Comprehensive Survey of Tools and Techniques

DEEPAK UNIYAL, Indian Institute of Technology Roorkee
VASKAR RAYCHOUDHURY, Indian Institute of Technology Roorkee

Pervasive healthcare is an emerging technology that aims to provide round-the-clock monitoring of several vital signs of patients using various health sensors, specialized communication protocols, and intelligent context-aware applications. Pervasive healthcare applications proactively contact the caregiver provided any abnormality arises in the health condition of a monitored patient. It has been a boon to the patients suffering from different diseases and requiring continuous monitoring and care, such as, disabled individuals, elderly and weak persons living alone, children of different ages, and adults who are susceptible to near-fatal falls or sudden increases in blood pressure, heart rates, stress level, etc. Existing surveys on pervasive healthcare cover generic techniques or a particular application, like fall detection. In this paper, we carry out a comprehensive coverage of several common disorders addressed by pervasive healthcare in recent years. We roughly classify different diseases by age groups of patients and then discuss various hardware and software tools and techniques to detect or treat them. We have also included detailed tabular classification of a large selection of significant research articles in pervasive healthcare.

## 1. INTRODUCTION

Healthcare is an industry which grows rapidly all over the world with an extreme pace. Both social and commercial implications of healthcare are manifold. Traditional healthcare services have two major drawbacks. Firstly, they are not available all the time and everywhere. Ailing individuals have to visit the caregivers or vice versa in order to start the treatment. This puts a constraint on elderly and/or disabled people living alone and requiring sudden medical attention to thwart possible long term handicap. Secondly, prevailing healthcare infrastructure and personnel are insufficient to cater to the needs of the increasing population. Moreover, the changing demographics, like the rapidly aging population in most parts of the world, and factors like pollution and stress put considerable strain on the already fragile healthcare infrastructure.

In order to address the above challenges, and to ensure maximum coverage, quality, and efficiency of healthcare services to everyone, everywhere, and all the time, pervasive healthcare solutions have been proposed which uses wearable sensors, wireless communications, and mobile computing to achieve its objectives. Rapid growth of sensor enabled smartphones has increased the outreach of pervasive healthcare. Today most of the mobile phones have a number of specialized sensors attached to them including accelerometer, proximity sensor, gyroscope, microphone, camera, light sensor, barometer, magnetometer, GPS, etc. Smartphones can be used to capture vital signs of users and alarm them accordingly if any aberrations are noted [Khan et al. 2012]. In case of serious injuries, caregivers can reach the patient by tracking his GPS coordinates. Sensor data collected over longer time period is analyzed through intelligent techniques to find out disorders which are not readily observable during the usually short patient-doctor interactions. Also, manual errors in vital signs or medical test data analysis can be reduced by using automated techniques.

Pervasive healthcare ensures remote monitoring of older individuals, especially those with Alzheimer's disease, Parkinson's disease, stroke patients in their rehabilitation phase, etc., who need continuous and ongoing medical support and care. Ambient sensors, like video cameras or image based sensors can be used to track the indoor activities and can raise alarm for unusual activities, like fall [Anderson et al. 2006][Schulze et al. 2009][Smith and Bagley 2010]. Video sensor based systems are also developed to help physically challenged people who suffer from the severe disabilities like amyotrophic lateral sclerosis (ALS) or cerebral palsy in which user have ability to move his eyes or head only [Ishimatsu et al. 1997][Krejcar 2011][L.J.G. et al. 2011][Su et al. 2008][Takami et al. 1996].

In this paper, we carry out a survey of pervasive healthcare techniques along with the related tools and technologies. There are several existing surveys [Acampora et al. 2013][Alemdar and Ersoy 2010][Pantelopoulos and Bourbakis 2010][Varshney 2007][Orwat et al. 2008] on using pervasive computing technologies in healthcare. [Acampora et al. 2013] discusses the infrastructure and enabling technologies required for applying Ambient Intelligence (AmI) techniques in the healthcare domain. They also summarize a set of algorithms and methods, like activity recognition, behavioral pattern discovery, decision support systems, etc. They finally include a set of existing applications of AmI, such as, continuous monitoring, smart hospitals, therapy and rehabilitation, assisted living, etc. [Alemdar and Ersoy 2010] discusses an unified architectural model and different wireless networking technologies followed by various pervasive healthcare applications. They studied the use of WSN in different healthcare applications, such as, fall and activity detection, location tracking, vital signs and medication intake monitoring, etc. [Pantelopoulos and Bourbakis 2010] reviewed several existing wearable bio-sensor systems and identified different qualitative and quantitative features, such as, wearability (light weight and small size), cost, etc. Then they assign some weight to the feature signifying the relative importance of the feature. Based on that, they classified the existing systems. [Varshney 2007] focuses on applications and requirements of pervasive healthcare, health monitoring using wireless LANs and ad hoc wireless networks. Their paper is focused more towards the wireless health monitoring technologies and less towards healthcare techniques, like disease detection and support systems. [Orwat et al. 2008] studies several pervasive healthcare research articles published during 2002-2006 and identified 69 articles describing 67 different systems. These articles have been categorized into different classes, based on status of the system (prototype, clinical trials, etc), users of the system (patients, care givers), healthcare focus (cardiovascular, respiratory, neurological, etc), and so on and so forth. The paper does not describe individual systems, but focus on a generic high-level classification of the systems. We, on the other hand, extensively study pervasive healthcare techniques adopted in detecting, monitoring, and treating various prominent health issues including disability, cardio vascular and stress related disorders, autism, cancer, obesity, falls, and Parkinson's disease. Our paper has been organized focusing on diseases or disorders mostly attributed to different age groups, like children, elderly, and adults. We also focus on health issues related especially to adult women and disabled individuals.

## 2. ENABLING TOOLS AND TECHNOLOGIES

Pervasive healthcare technique works by a reliable coordination of multi-sensor sensing, short range wireless communications, and data mining based decision making strategies [Seyed Amir Hoseini-tabatabaei et al. 2013]. Here we give brief accounts of various sensing, communication, and decision making techniques used in pervasive healthcare.

### 2.1 Sensor Technology

Pervasive Healthcare aims to continuously monitor several vital signs of patients using different health sensors either embedded in some device, like smartphone, or wearable by users. Sensors continuously monitor the patient and transfer the data to the server. Most of the existing research works in pervasive healthcare makes use of one or more of the sensors listed in Table 2. Ideally, the sensor we choose should satisfy the following requirements [Wilson and Atkeson 2005].

- Invisible, unobtrusive and non-invasive – should not interfere in user's daily activities
- Protect user privacy and data confidentiality
- Consume low-power and demand minimal computational resources
- Inexpensive and easy to install
- Low-maintenance overhead

Different research works may use the same sensor to address different problems. E.g., accelerometer can be used for fall detection, child care, disabled care, or for eating gesture recognition is obesity control. A summary of the healthcare problems addressed by different sensors have been provided in Table 3.

### 2.2 Communication Technologies

Short-range wireless communication techniques are used to interconnect wearable health sensors in order to form a Body Area Network (BAN). Transferring sensor data the central server, however, needs long range wired or wireless communication techniques. Table 4 summarizes the existing communication technologies used in pervasive healthcare applications along with their features. RFID is a novel communication technology which works using radio frequency.

### 2.3 Decision Making Techniques

Several standard data mining techniques are used for intelligent decision making. Sensor data is classified using Decision Trees, Artificial Neural Networks (ANN), Fuzzy Logic, Hidden Markov Model (HMM) and its variants, Support Vector Machine (SVM), Bayesian Classifiers, Bayesian Networks, clustering techniques, like k Nearest Neighbor (kNN) and k-Means.

## 3. CLASSIFICATION OF DISEASES

Associating diseases with a particular age group or age or any gender is difficult due to the changing nature of diseases and the patients. With the increased stress level in our daily life, any disease can strike us at any time and without notice. The objective of pervasive healthcare is to keep track of the symptoms of a disease creeping up in our body and to inform or warn us beforehand. Due to the want of a better way to classify diseases, in this paper, we resorted to the usual age and gender based classifications as shown in the Figure 1 below. The major classes are elderly healthcare, child healthcare, adult healthcare, women healthcare, and disabled care.

Elderly healthcare revolves around the central idea of *fall detection* and *prevention*. Falls are major health risks among elderly that diminish the quality of life and are major causes of pain, disability, loss of independence, and premature death. According to WHO global report on "falls prevention in older age", fatal fall rates increase exponentially with age and frailty level for both sexes, highest at the age of 85 years and over. Another disease usually associated with the old age is the Parkinson's disease (PD). Average age of the disease onset is 55 but about 10% cases are in younger than 40 years old. So, in this paper, under elderly healthcare, we actually surveyed multiple fall detection, prevention, and analysis techniques as well as the detection and treatment methods of PD.

Child healthcare is the next major class of healthcare service provisioning which considers various stages of pre and postnatal cares. Prenatal care includes foetal care, like foetal HR monitoring, and postnatal care includes care for diseases like Autism, Cerebral palsy, etc.

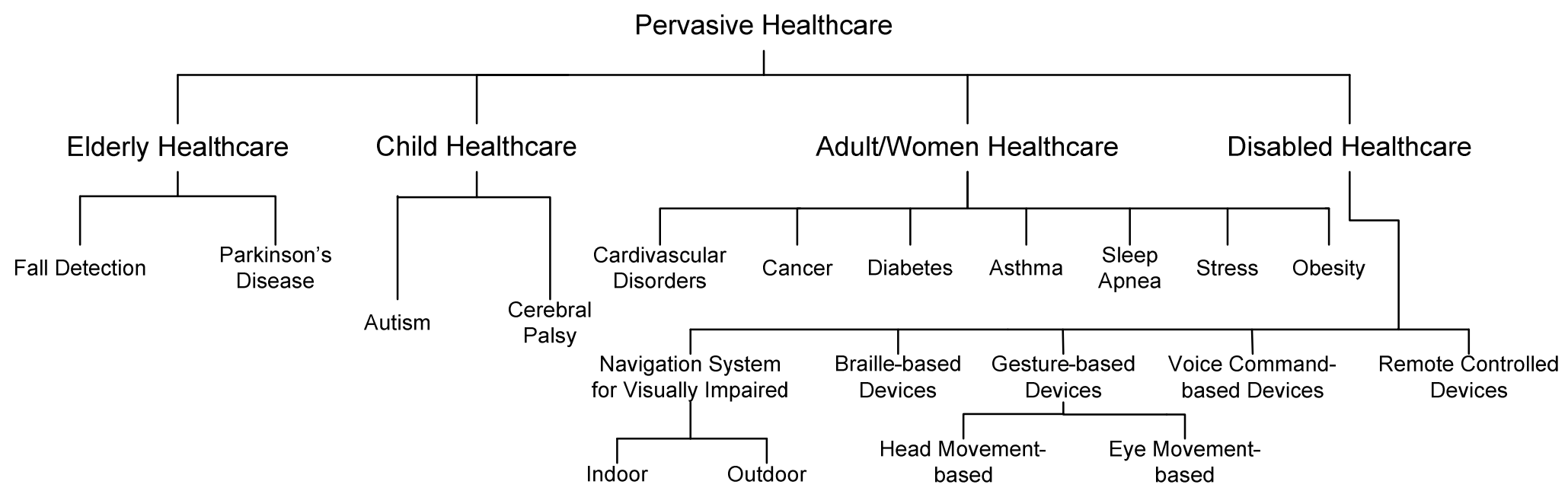

**Fig. 1: Classification of Pervasive Healthcare Domains**

*Adult healthcare* focuses on a medley of diseases none of which can be exactly associated with a particular age or gender and can take place at any time of the adult life and in many cases even in the childhood. They include *cardiovascular* disorders, *diabetes*, *asthma*, *sleep apnea*, *stress*, and *obesity*. While the last two are not exactly diseases, they are disease-like conditions which pave the way for more severe health disorders. Under *women healthcare* we have discussed many women specific health conditions including pregnancy and foetal health monitoring techniques. Since, all the predominant health conditions for adult women are discussed under the adult healthcare, we have merged them into a common category titled *adult/women healthcare*.

Disabled healthcare discusses multiple techniques to enable individuals with disabilities to interact with the environment as smoothly as possible. Below we discuss different classes of healthcare services in more details.

## 4. ELDERLY HEALTHCARE

Elderly population is growing rapidly all over the world. They are at a high risk of developing various chronic illnesses (e.g., diabetes, arthritis, stroke, Ischaemic heart disease, etc) which result in several disabilities and sometimes to death. Pervasive healthcare can be used in these cases to monitor vital signs, like blood pressure (BP), blood glucose level and to take action accordingly. We have discussed detection and treatment techniques of several such diseases under adult healthcare as they are not specific to old age particularly. However, tremendous amount of research in elderly healthcare have been carried out to detect accidental falls which is a major health risk among elderly and to treat/assist people with PD. Either of these health conditions may render elderly individuals incapable of living alone. In Sections 4.1 and 4.2 respectively, we discuss in-depth several methods of detecting and treating falls and Parkinson's disease cases using pervasive healthcare technology.

### 4.1 Fall Detection

Falls are the biggest threat among all events to the elderly people and patients [Yu 2008] and poses a big risk to well-being, confidence, and mortality of elderly [Lord and Colvin 1991]**.** Statistics [Griffiths et al. 2005] show that falls are leading reason for injury related death for senior citizen aged 79 or more and second leading cause of injury related death for all ages. With the rapid growth of elderly population, demand of intelligent systems has been increased in the healthcare industry to promptly detect falls and to communicate it to the care givers accordingly. It has been proved that medical consequences of a fall are highly dependent on the rescue and response time. So, the objective of all fall detection systems (FDS) (see Table 5) is to minimize the delay in detecting a real fall to alarm the emergency services accordingly. Activity recognition [Wilson and Atkeson 2005], dementia travelling [Lin et al. 2012] presents systems that can provide safety and health related services to elderly by recognizing their daily activities and their behavior.

Fall detection is challenging because it is difficult to distinguish between real falls and some other activities of daily living (ADL) or simple object falling events. Also, falls are of different types [Yu 2008], like falls during sleeping (from bed), falls during sitting (from chair), falls during walking and standing on the floor, and falls from standing on some support, such as ladder or some stool.

Falls can be detected using both *hardware-based* and *software based* methods. Hardware based methods are those which use one or more assistive tools for uniquely identifying true falls in real-time using acceleration, body posture, and other useful parameters. Software based methods, on the other hand, apply widely available data mining techniques on pre-stored sensor data readings to identify falls from other daily activities.

Fig. 2 shows afore-mentioned classification of several FDS. In-depth discussion of several hardware and software based fall detection mechanisms along with their characteristic features shall be discussed in Section 4.1.1 and Section 4.1.2, respectively.

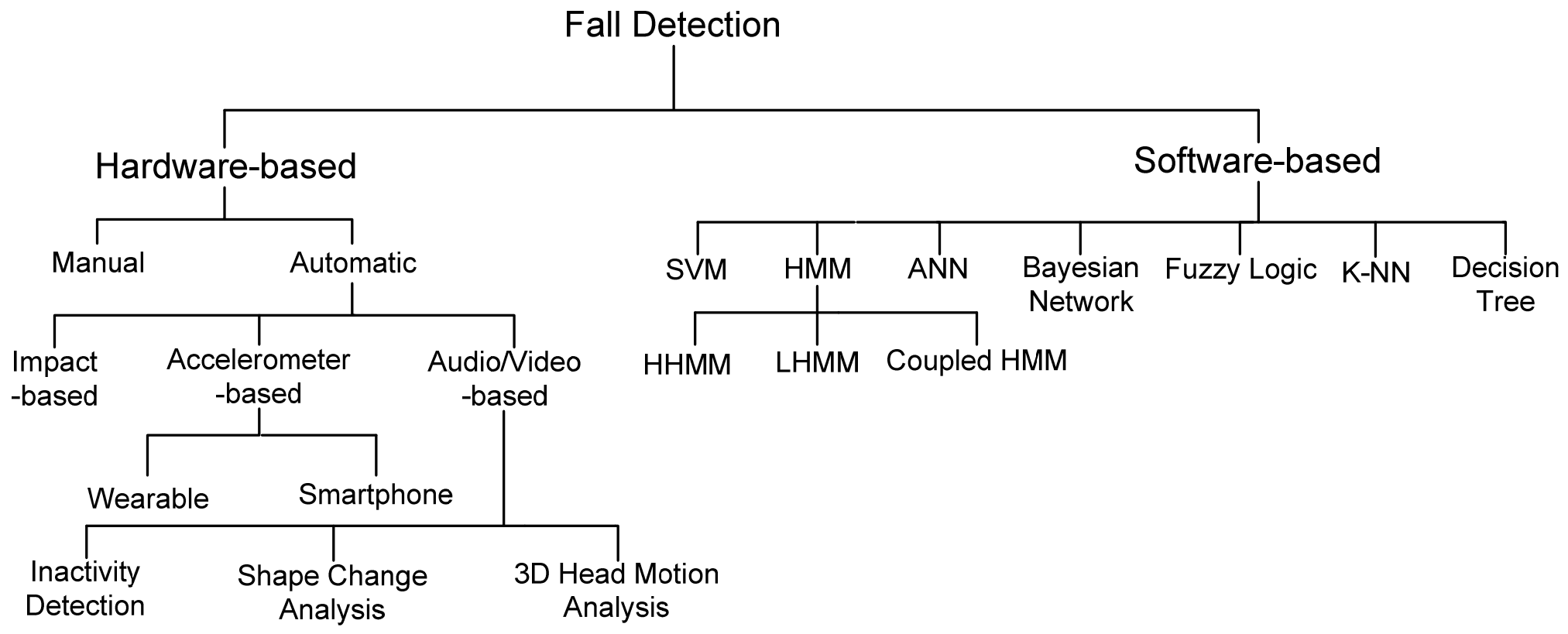

**Fig. 2: Classification of Fall Detection Techniques**

*4.1.1 Hardware-based Fall Detection.* Hardware-based fall detection can be both *manual* and *automatic*. The manual fall detector works trivially by assuming that the person suffering from the fall shall voluntarily inform about his fall through some device, like pressing-a-button. However, this might not be possible always given the seriousness of the fall and the sense-state of the fallen individual. So, a more appropriate though difficult way is to detect falls automatically. Automatic fall detection can be of three types. The first one is *post-impact* fall detection which is an easier way given the difficulty of distinguishing falls from many other ADLs. Second type of automatic fall detection method uses several augmented wearable sensing devices as well as smartphones. The third and final type monitors ADL through audio/video recordings and then captures fall events from those streams. Below we discuss different types of fall detection techniques in more detail.

*4.1.1.1. Manual/ User Activated Fall Detection.* User activated fall detection [Alwan et al. 2006] devices are RFID-enabled wrist bands equipped with a push-button which can be pressed when a user is falling or has fallen. The RFID reader on receiving the user signal, communicates with the emergency services to expedite the relief actions. This type of system is easy to use, low cost, and simple but they have limited usability for more debilitating falls leading to serious injury and/or loss of consciousness. Moreover, user activation may fail and may not be effective in cases where the user have memory disorders like Alzheimer's disease, amnesia, agnosia, dementia,

Parkinson's disease etc. Also, user may not be wearing the device when a fall occurred against the implicit assumption that the user is always wearing the device. This type of system can only be used in indoor environments. RFID communication technology used for these systems [Chen and Wang 2007] is free from the line-of-sight problem. However, it works only in a limited range of 20-100 meters.

*4.1.1.2. Automatic Fall Detection.* Various automatic FDS have been designed which do not require manual activation. These systems may be *impact based*, *accelerometer based* or *audio/video based* which trigger the alarm automatically when it detects some fall. Automatic devices require some previous setup and the user need not bother about manually involve in initiating communication like in user activated systems. Automatic devices collect sensor data and detect fall detection based on some algorithms developed. This survey[Hijaz et al. 2010] conducts a survey on various automatic techniques and algorithms developed for detecting fall or any abnormality in daily activities. After an extensive study it defines the different techniques in three main categories: (1) Video Analysis Based; (2) Acoustic and Ambience Sensor Based; and (3) Kinematic Sensor Based.

*Impact (Vibration/ Acoustic) Based.* Impact-based fall detection is facilitated with the use of pressure sensors embedded in the floor mat. When a person falls, the resulting vibration is recorded and a fall detection mechanism is triggered. This kind of system works on the presumption that the vibration pattern generated by human fall is different than that produced by other ADLs and normal object falling [Alwan et al. 2006]. Another such work [Alwan et al. 2006] uses piezoelectric sensor embedded in the floor surface along with some mass spring arrangement and electronic device to detect the fall pattern. After detecting a fall they use telephone line to communicate with the caregiver. The accuracy reported is fairly high compared to previous works. An acoustic FDS has been introduced in [Zhuang et al. 2009] which uses input from a single microphone and models each fall or noise with Gaussian mixture model (GMM) and employ SVM on its kernel to distinguish between noise and real falls.

*Accelerometer-based.* Research works [Perry et al. 2009] have been carried out to show that the fall detection methods using accelerometer are more accurate and reliable compared to other methods. Accelerometers used are either *embedded in smartphone* [Sposaro and Tyson 2009] or they are *wearable* - worn in wrist, waist, trunk, chest, thigh, or even kept in pant pockets [Bourke et al. 2007]. Below we discuss the two types in more details. Earlier accelerometers used to be uni-axial but recent advancements in MEMS technology has led to the development of bi-axial and tri-axial accelerometers (TAA). One such system [Selvabala and Ganesh 2012] for fall detection uses a low power MEMS based accelerometer and Passive Infrared (PIR) sensor. Falls are detected based on voltage changes due to human activities. The system can distinguish between real falls of human being and objects of same weight with an accuracy of 99%.

— *Smartphone-embedded.* Smartphone embedded accelerometers and sometimes other external sensing entities attached to it are used to detect falls. One such system has been proposed in [Abbate et al. 2012] which uses HTC Google Nexus One smartphone (with an embedded accelerometer) and another external sensing unit with accelerometer. A fall alarm is raised only when an acceleration peak greater than 3g is followed by a period of 2500 ms during which no further peaks exceeding the threshold takes place. The accelerometer sampling rate has been set at 50 Hz. The 3g value is small enough to avoid false negatives so this paper discusses different techniques to avoid false alarms. A neural network based pattern classifier has been developed to distinguish real falls from sitting/lying,

jumping/running/ walking, and hitting the sensor. PerFallD [Dai et al. 2010] is a Android G1 phone-based fall detection mechanism which uses two different threshold acceleration values. If the pre-specified condition about both acceleration values are satisfied fall is said to be detected. The smartphone can be attached with chest, waist, or thigh (pants' pocket), though placing it on the waist performs best with least false negatives and false positives. FallAlarm [Zhao et al. 2012] is a Nokia N95 smartphone based FDS which uses decision tree to analyze the data taken from the embedded accelerometer for detecting fall. Moreover, the device can also detect the location of the fall based on Wi-Fi signals.

— *Wearable*. There are various wearable accelerometer based FDS. [Karantonis et al. 2006] proposed a waist-mounted TAA based system to distinguish between period of activity, rest, posture, orientation, and events like walking and fall. Another TAA based FDS [Bourke et al. 2007] worn on trunk or thigh uses pre-defined upper (trunk) and lower (thigh) thresholds to distinguish between ADL and real falls. Another trunk-mounted TAA based system [Jantaraprim et al. 2010] performs better fall detection using free fall characteristics based on 'free fall' and 'beginning to maximum peak'. Another accelerometer-based wrist-worn interactive device aims to prevent falls by continuously monitoring the user and providing feedbacks about actual risk. So this fall prevention system make user aware of adverse effects of falling. Experiments of fall-detection using a sensor-embedded vest [Bourke et al. 2008] worn under clothing has also been carried out. A hybrid of TAA and video camera based activity detection device called SmartBuckle [Cho et al. 2008] is mounted to belt and can detect human activities with an overall accuracy rate of 93%. It then extracts the features of different activities and classifies them using SVM. One such activity can be fall activity. Another system [Noury et al. 2000] with three different embedded sensors use piezoelectric accelerometer for vertical acceleration shock, position tilts switch for body orientation, and sensor for mechanical vibrations of the body. There are also several commercially available FDS, like LifeAlert. Wearable sensor-based fall detectors are easy to install and setup, easy to operate, and are cost-effective. However, they intrude upon user privacy [Mubashir et al. 2013] and may create discomfort. Also, user may forget [Lai et al. 2010][Fang et al. 2012] to wear them at all times. Fall detection may not be easy in these types of systems as it may depend on the position of sensor in the human body.

*Audio/Video Based.* Audio/Video based techniques for fall detection uses video camera [Anderson et al. 2006][Schulze et al. 2009][Smith and Bagley 2010][Hsiao et al. 2011][Lin and Ling 2007] to monitor people's ADLs and can detect falls using some threshold techniques. They are more advantageous than wearable device based techniques because they are less intrusive, can detect multiple events simultaneously, and also the recorded video can be used for remote and post verification and analysis. However, video camera surveillance is not available at every place and they suffer from serious privacy concerns. Video camera based fall detection techniques may be costly due to the requirement of large and expensive cameras to cover every corner of a house. One approach to use extremely low power cameras to automatically detect and localize falls has been proposed in [Williams et al. 2007]. Low power cameras are attractive choice but they are not computationally efficient to execute complex computer vision algorithms [Nait-Charif and McKenna 2004][Töreyin et al. 2005]. Privacy concern of using a video camera has been checked in [Anderson et al. 2006] by using silhouettes instead of real human beings. It captures live video at 3FPS and extracts silhouettes based on statistically modeling a static background and then segmenting the human based on color information. After feature extraction from silhouette HMM is trained for temporal pattern recognition and it is used in future for recognizing similar activities. There are other works [Chen, He, Keller, et al. 2006][Chen, He, Anderson, et al. 2006] for activity recognition based on silhouette extraction. Other systems which use camera based fall detection are [Schulze et al. 2009][Smith and Bagley 2010] [Hsiao et al. 2011][Lin and Ling 2007] [Hauptmann et al. 2003][Williams et al. 2007][Töreyin et al. 2005] [Diraco et al. 2010][Abu-Faraj et al. 2010][Thome et al. 2008][Liu et al. 2010][Oliver et al. 2000][Brand and Kettnaker 2000]. Vision based FDS have been classified in [Diraco et al. 2010] into 3 types - *inactivity detection*, *shape change analysis*, and *3D head motion analysis*.

—*Inactivity detection.* It is based on the principle that fall will end with an activity period on the floor and head lies on the floor motionless or with very less motion at the end of process. [Nait-Charif and McKenna 2004] uses low cost wide angle cameras mounted on ceiling. [Jansen and Deklerck 2006] Extracts information from images obtained using 3D camera technology and combines that orientation information of the body with inactivity detection.

—*Shape change analysis.* Shape of person will change from standing to lying on the floor. The head lies on the floor at the end of fall with no or little motion. [Shieh and Huang 2009] Uses multiple cameras to fetch the images from different regions and then use falling pattern recognition approach to detect fall.

—*3D head motion analysis.* Based on the principle that vertical motion is faster than horizontal motion during fall and a fall is detected using 3D head velocities of head computed from 3D head trajectory. Head is chosen for tracking [Rougier and Meunier 2005][Rougier et al. 2007][Rougier et al. 2006] as it is visible most of the time and has a large movement during fall.

*4.1.2 Software-based Fall Detection.* There are many accelerometer-based systems which apply data mining techniques over sensor-collected data to aid the detection of real falls. They mostly classify ADLs including falls using different data mining techniques as described below.

*4.1.2.1. Support vector Machine (SVM).* Several research works [Lai et al. 2010][Cho et al. 2008] [Hauptmann et al. 2003] use SVM to classify daily user activities. Williams et al. [Williams et al. 2007], uses SVM to classify the activities as fall or no fall. Sometimes balance impairments in elderly may lead to falls. Feature extraction and classification using SVM has been carried out in [Khandoker et al. 2007] for automated screening of gait patterns related to balance impairments. Zhuang et al. [Zhuang et al. 2009] presents an acoustic FDS which uses input from a single microphone and models each fall or noise with GMM and employ SVM on its kernel to distinguish between noise and real falls.

*4.1.2.2. Hidden Markov Model (HMM).* There are several research works which model human motion using data collected from sensor nodes and following standard HMM [Anderson et al. 2006][Töreyin et al. 2005][Oliver et al. 2000][Kasteren et al. 2008][Thome and Miguet 2006] and its different variants, such as, Entropic-HMMs [Brand and Kettnaker 2000], coupled-HMMs (CHMM) [Oliver et al. 2000][Brand et al. 1997], and Layered-HMM [Thome et al. 2008] (a special case of Hierarchical HMM).

*4.1.2.3. Artificial Neural network.* [Sixsmith and Johnson 2004] uses Neural Network to classify falls using vertical velocity estimates derived from Infra-Red Integrated Systems (IRISYS) sensor data.

*4.1.2.4. Bayesian Network.* [Diraco et al. 2010] Bayesian segmentation is used to extract quantitative information about the subject under consideration and determine the distance of its center-of-mass from the floor plane.

*4.1.2.5. Fuzzy Logic.* [Chen, He, Keller, et al. 2006][Chen, He, Anderson, et al. 2006] uses fuzzy logic to detach the moving object from human silhouette. Fuzzy logic is used in [Putchana et al. 2012] for movement analysis (over accelerometer data) to keep the computational cost low. The system can distinguish between different movements, like standing, sitting, forward fall, and backward fall. Highest accuracy for detecting standing position is 99.56% and for forward fall it is 96.34%. However, the accuracy drops to 89.26% for sitting and to 77.49% for backward fall.

*4.1.2.6. K-nearest Neighbor (K-NN).* Value of k depends on the data. Higher value may reduce the effect of noise on classification but boundaries of different classes will not be much distinguishable [Liu et al. 2010] uses K-NN classification scheme to perform posture classification using ratio and difference of human body silhouettes. K-NN is used in [Ghasemzadeh et al. 2010] over sensor data collected through TAA, bi-axial gyroscope, and an webcam to capture human movements for classification of activities. The system achieves a classification accuracy of 84.13% for different activities such as standing, sitting, bending, moving forward, moving backward, etc.

*4.1.2.7. Decision Tree.* [Ermes et al. 2008] uses custom decision tree, automatically generated decision tree, ANN and Hybrid model for activity classification. Activity recognition was performed on features like mean, energy, entropy and correlation extracted from accelerometer data [Bao and Intille 2004] using decision table, instance based learning (IBL or nearest neighbor), decision tree and naive Bayes classifiers. Decision tree gave best result (accuracy over 80%) followed by nearest neighbor, Bayes, and decision table.

### 4.2 Parkinson’s Disease

Parkinson's disease (PD) is a slowly progressive neurological disorder that affects movement, muscle control, and balance. PD was first described in 1817 by James Parkinson, a British physician who published a paper on what he called the shaking palsy. The exact cause of the disease is unknown, but scientists think that a combination of genetic and environmental factors can trigger it. Average age of the disease onset is 55 but about 10% of the patients are younger than 40 years old. There are rare chances of PD appearing in people before the age of 20, and the condition is called *Juvenile Parkinsonism* which is commonly found in Japan. PD is not fatal but can lead to severe incapacity within 10-20 years. It is estimated that 90% of people with PD suffer from speech and voice disorders. Vocal cords are damaged by this disease and creates an improper voice in the patient’s speech [Shirvan and Tahami 2011]. So, various features of voice or speech can be extracted and analyzed to predict severity of PD.

The Unified Parkinson's Disease Rating Scale (UPDRS) is the most reliable and widely used clinical rating scale (ranging from 0 (normal) – 4 (severe)) to check the severity of PD. However, several key elements are not covered under UPDRS in order to make the scale reasonably simple and short. Moreover, some of the metric characteristics that support the validity of a tool have never been checked [Martinez-Martin et al. 1994] for the UPDRS. Using UPDRS requires a trained Neurologist available only in the clinical environment. The scale is qualitatively interpreted, so there may be variation among multiple Neurologists. Also, the scale is ordinal, thus lacking a temporal parameter to evaluate the attributes of disorder in movement. Tremor characteristics and temporal attributes of PD can be readily quantified using accelerometer [LeMoyne et al. 2009]. Research works developing techniques to detect and to treat PD, and to assist PD patients have been summarized in Table 6.

*4.2.1 Quantitative Methods to Detect PD.* There are following four methods to detect PD.

*4.2.1.1. Finger-tapping Method.* Finger tapping based PD detection method [Okuno et al. 2007][Okuno et al. 2006] uses TAA and touch sensors (made of thin stainless sheets) to be worn on the index finger and the thumb of the patient. A transfer function was designed which gives a relation between accelerometer readings and contact force between index finger and thumb. The collected data is evaluated by a neurologist and the UPDRS score for finger tap is calculated. The whole process shows that the contact force decreases with increasing score of UPDRS finger tap test and therefore finger tapping could be an important factor in determining the severity of PD. There are some systems [Shima et al. 2008][Kandori et al. 2004] which improves on finger tapping method by using 2 magnetic sensor coils attached to the distal parts of fingers. The output voltage is based on the distance between fingertips and is stored in a PC. Experimental results indicate that PD patients show larger variation in tapping rhythm than normal subjects. There are also some other finger tapping based PD detection systems [Stamatakis et al. 2013][Agostino et al. 2003][Konczak et al. 1997][Kupryjanow et al. 2010][Shima et al. 2009]. [Stamatakis et al. 2013] uses video camera to record finger tapping task and then scoring is done on the basis of MDS-UPDRS instructions [Goetz et al. 2008]. [Agostino et al. 2003] shows that compared to normal subjects PD patients are slow in finger flexion and switching between flexion and expansion. [Konczak et al. 1997] uses both finger tapping and lips movement performed simultaneously or in isolation for detecting the severity of PD. [Kupryjanow et al. 2010] uses a virtual-touchpad to capture the image of finger tapping and alternating rapid hand movements. SVM classifiers are used to distinguish between different such gestures and UPDRS test score is provided to detect the severity of PD. Fingerpad stiffness model [Shima et al. 2009] is developed to evaluate the relationship between finger pad forces and deformation caused during finger tapping movement. It is found that the forces generated by PD candidate differ from normal person.

*4.2.1.2. Speech-based Method.* PD often affects the coordination of different components for smooth and fluent speech [Asgari and Shafran 2010b]. Speech rates of PD patients differ from normal persons, and it may be faster or slower as compared to that of a normal person [Schulz and Grant 2000]. Voice intensity reduction, unvarying pitch, breathy harsh voice, monotony of speech, and an abnormal rate of speaking are the main features of speech disorder in PD subjects that reduce overall intelligibility of speech [Asgari and Shafran 2010a][Darley et al. 1969b][Darley et al. 1969a][Scott and Caird 1983]. Using the speech features like pitch, spectral entropy (differentiate between speech and noise), harmonic to noise ratio, etc. [Asgari and Shafran 2010a][Asgari and Shafran 2010b] performs three different tests for PD, like *sustained phonation* (phonate vowels in clear and steady voice as long as possible), *diadochokinetic* (DDK) (how quickly a person can accurately repeat a sequence of syllable containing consonant-vowel combination) and *reading*. Ratings of severity of PD are measured using the features extracted from speech during those tests with the help different regression models. Best performance occurs when regression model is estimated by Epsilon SVM with polynomial kernel of degree 3. Another system [Bocklet et al. 2011] also uses SVM classifier to differentiate between the normal speech and speech of PD candidate. Temporospatial analyses of oropharynx structure, specifically the velum, are performed [Robbins et al. 1986] on the data collected by Video fluoroscopy monitoring during swallowing and speech production. These activities are chosen because oropharynx system is prominently involved in both the process and result shows that all the PD patients show abnormal movements of oropharynx structure. This kind of monitoring can be very helpful in detecting and managing aspiration (inhaling fluid or stomach contents) at early stage. Different approaches to treat speech disorders caused by PD are discussed in [Schulz and Grant 2000]. It concludes that neither pharmacological nor surgical methods like thalamotomy, Pallidotomy, fetal cell transplantation (FCT), and deep brain simulation can alone treat and improve the speech function significantly in PD subjects. Currently speech therapy with pharmacological methods is proved to be most effective for improving voice functions.

*4.2.1.3. Spiral Analysis.* PD generally affects elderly people and can be recognized with four cardinal features - tremor, rigidity, slowness of movement, and postural instability [Jankovic 2008][Savitt et al. 2006][Surangsrirat and Thanawattano 2012]. Some computer based tools for diagnosis of PD depends upon those four features. Spiral analysis is a computerized method that tries to detect movement related dysfunction from a simple drawing of Archimedean spirals in repetitive manner [Wang et al. 2008]. Spiral analysis is based on the ability to draw straight line or circle either directly on a tablet or over a paper (which can be scanned and inserted in a computer). The drawings are then programmatically analyzed to calculate the mean and standard deviation of distances. This kind of analysis does not require any complex process. Many systems have been developed for spiral analysis. [Surangsrirat and Thanawattano 2012] uses touchscreen android OS tablet and asks user to trace Archimedean and octagon spirals with the help of stylus to detect severity of PD. [Dounskaia et al. 2009] asks user to draw on a digitizer-tablet with an inkless pen and the movements are recorded in orthogonal coordinate system. In [Miralles et al. 2006] user draws the spiral over a printed template which is later scanner with semi-automatic computer program. [Cunningham et al. 2009] uses a different strategy other than drawing line or circle as it asks user to click on the center of 10 different points shown in computer screen. Each point is divided into concentric circles and clicking the target on its outer radius or more deviation from the center shows the higher severity of PD. Unfortunately this deviation also occurs due to computer illiteracy and hence, it is difficult to determine the actual cause. [Westin et al. 2010] uses DWT and PCA to process the spiral drawings and then generates the test scores to quantitatively determine the impairment.

*4.2.1.4. Inertial Sensor (Accelerometer/Gyroscope) Based Systems.* A three-tier system named Mercury Live [B.-R. Chen et al. 2011] uses accelerometer to collect data and web application to remotely connect with an expert. Physician can ask user to perform some motor tasks and evaluate the UPDRS score on the basis of real time data collected though wearable inertial sensors. In another system [LeMoyne et al. 2010], iPhone containing inbuilt TAA is mounted on dorsum of hand through a glove. This system measures the severity of PD with the help of tremor attributes and sends collected data to a remote server for further processing and analysis. The preliminary evaluation and testing shows a positive result that tremor attributes can be successfully used for detecting PD while putting very less or no strain on highly specialized medical resources. In a system developed by [Hoffman and McNames 2011], user wears accelerometer and gyroscope on second phalanx of his index finger and three finger tapping exercises (pad-pad finger tap, tip-knuckle finger tap, and hand-pronation supination) are performed twice each. Collected data is send to a nearby laptop with the help of Bluetooth connection and later processed with MATLAB. Overall assessment shows that inertial sensors can be very promising for quantifying motor impairment. Simple test such as timed up and go (TUG) is used to assess the mobility of a person but it’s result is dependent upon the time measured by stopwatch and that is why it is not practical and easy-to-use. So to eliminate the limitations of existing methods, [Mariani et al. 2013] uses inertial sensors that are worn on shoes to analyze simple tests like timed up and go (TUG) and long distance walking which is practical and easier to use.

*4.2.2 Treatments for PD.* Several different techniques, such as, Drug Therapy, Pallidotomy, and Deep brain simulation (DBS) [Nolte and Sundsten 2002][Kandel et al. 2000][Volkmann et al. 2006] have been developed to treat PD. However, since, PD is caused by lack of release or loss of dopamine (an important chemical in brain), there is no complete cure for it. Medication and surgery can only provide significant relief from the symptoms .

*4.2.2.1. Speech Therapy.* Difficulty in speaking and swallowing are severely limiting symptoms of PD but speech therapy can be helpful in diagnosing both. Experts teach the patients communication skills that may be verbal or non-verbal and thus helping the person in better communication. Some research focused on the devices which include voice amplifier, etc. Voice amplifier increases loudness of voice thus relieving any anxiety a PD patient may feel [Schulz and Grant 2000].

*4.2.2.2. Pharmacological.* Medication for PD falls under three categories. First one includes drugs that work to increase the level of dopamine in brain; second one includes drug that affects some neurotransmitters in the body to soothe PD symptoms, while the third one includes medications that help in controlling non-motor functions that do not affect movement.

*4.2.2.3. Surgical Treatments:* There are following surgical treatments available for PD.

*Thalamotomy and Pallidotomy.* They are the earliest types of surgery and they try to reduce the PD symptoms by destroying specific sections of brain. Pallidotomy works by removing a portion of the brain called the *globus pallidus*, while part of the brain's thalamus is destroyed in Thalamotomy. Although these surgical methods can improve the symptoms of tremor, they lead to permanent destruction of brain tissues and therefore these procedures are replaced by deep brain simulation [Schulz and Grant 2000].

*Fetal Cell Transplantation (FCT).* FCT refers to the replacement of dopamine releasing cells exactly where they are needed, in the striatum.

*Deep Brain Simulation (DBS).* DBS uses a medical device called brain pacemaker which sends electrical impulses to specific parts of the brain. DBS can be classified as *open loop* and *closed loop* DBS. Open loop cannot automatically adapt the simulation according to the patient's condition, and continuous high frequency electrical simulation delivered to the brain would cause side effects to the patient. Therefore closed loop DBS is a must for offering appropriate amount of electric simulation to the brain [Hu et al. 2013]. The mechanism of DBS is still unclear and this lack of understanding makes the selection of simulation parameters (i.e. voltage, pulse duration, and frequency) quite challenging [Santaniello et al. 2011][Rosin et al. 2011].

## 5. CHILD HEALTHCARE

Pervasive healthcare for children uses several technological solutions or supports to children in need. In 2011 Centers for Disease Control and Prevention (CDC) reported that during 1980-2008 obesity rates tripled for children. Obesity can lead to a number of diseases including high BP, diabetes, heart disease, joint problems, like osteoarthritis, sleep apnea, and respiratory problems, cancer, metabolic syndrome, psychological effects, etc.

Cerebral palsy is another child-specific disorder which can be defined as a group of non-progressive, non-contagious motor conditions that cause physical disability. Children with cerebral palsy may have difficulty in body movement and may fall frequently. System [Smith and Bagley 2010] have been developed to help children with cerebral palsy, which is kept in child's fanny pack in lower back for analyzing and assessing the impact of treatment on daily activities of child so that better medication can be adopted according to the child's performance.

Many other pervasive healthcare solutions have been developed to help children suffering from chronic pain [Cann et al. 2012][Starida et al. 2003], autism [Chuah and DiBlasio 2012][Matson et al. 2008][Twyman et al. 2009][Roberts et al. 2011], and hypertension [Pruette et al. 2013]. Also, remote HR acquisition system [Marques et al. 2000], training system for child minders [Fujiwara et al. 2011], intelligent system for maternal and child-care [L. Chen et al. 2011], helps to provide better care to children suffering from different disease conditions. In Table 7 we have summarized various child healthcare systems. In the following subsection, we discuss pervasive healthcare solutions for autism in greater details.

### 5.1 Autism

Autism is a group of lifelong complex disorders of brain development, collectively called Autism Spectrum Disorder (ASD) [Yates and Couteur 2013]. The term "spectrum" refers to the wide range of symptoms, skills, and levels of impairment, or disability, that children with ASD can have. These disorders are characterized by difficulties in social interaction, verbal and non-verbal communication. Exact causes of ASD are not known but according to research both genes and environment play important roles.

50% of parents report autism features in their child by the age of 2 but 93% are able to recognize symptoms by age of 3. So the diagnosis age for autism is around 3-4 years before which it is hard to detect. A most recently adopted strategy is to record the video of daily activities of infants and let it be monitored by some expert to judge abnormal behavior or stereotypical activity. No test has been developed yet that can be performed to extend the capability of existing techniques to diagnose the ASD before age 2 [Matson et al. 2008].

There may be several other reasons behind undetected autism for very long time such as child is not given much chance for peer interaction in society or school and there is no standard instrument to easily monitor the child [Robins et al. 2001].

It is shown that there is much time difference between when parents are first concerned, first medical evaluation, and the age of first confirmed diagnosis. So it indicates that early intervention program may be very effective in well-being of child and his different activities in society. Early enrolment programs can address the speech, social behavior, and help the child to achieve educational and other goals [Twyman et al. 2009][Roberts et al. 2011].

Many systems have been developed to help autistic children. Stereotypical behaviors, such as body-rocking, hand waving, walking, jumping, etc. are exhibited by the children having ASD. The appropriate tools for behavior measurement are not either easily available or expensive to deploy. There are no systems available that can monitor the autistic children reliably and accurately. So the easy and low cost method for children behavior monitoring is performed by caregivers on their observations in classroom or at home. One such basic and simple method is paper-pencil recordings in which stereotypical activities of autistic child are recorded by caregivers in home or school [Chuah and DiBlasio 2012]. But this method may not detect stereotypical behavior with full accuracy and also depend on expertise of caregivers to monitor and analyze the abnormal behavior. Data storage and sharing is also a bigger concern in this kind of approach [Ellertson 2012].

The shortage of trained caregivers can be addressed by another method which involves video recording of behaviors and annotation of stereotypical activities by experts off-line on the server side. This method has an advantage that videos can be replayed and analyzed later in case of any question. This method is obviously more reliable and uniform than paper-pencil method but still very expensive, tedious, and time consuming [Sturmey 2003].

Both the methods mentioned above require one or more dedicated experts to analyze the child. To do away with this, some automatic systems have been proposed (as below) that can save cost as well as human effort.

[Westeyn et al. 2005] presents an on-body sensing system for monitoring stimming (repetitive body movement) activities. 68.57% accuracy is achieved by using HMM which distinguishes between stimming and non-stimming behavior. But this system does not analyze real data as it contains the accelerometer data generated by mimicking the stereotypical behavior of autistic children. Also this data does not contain any sensor data that describe the autistic children's environmental contexts.

Smartphone based systems for Autism social alert have been proposed in [Chuah and DiBlasio 2012] and [Ellertson 2012] which employs several sensors to monitor children activities and also environmental context to remove the drawback of previous systems. GPS, Bluetooth, audio readings are helpful in providing information related to environment and TAA provides the information related to child movements. Video recording is also performed so that any unanswered question can be resolved easily. J.48 classifier from the WEKA package is used to classify the different activities using the data collected from various sensors and the triggering factor of particular activity can be analyzed with the help of recorded environmental data. The immediate action can be taken to give a child positive reminder if the factor behind that activity is well known [Chuah and DiBlasio 2012].

Many therapies for autism have been developed in recent past, such as drug therapy, psycho-education, behavioral therapy, music therapy, etc. These therapies can be beneficial up to some extent but most of them are costly, abstract, complicated and takes a long time for treatment as well as learning the therapy. Compared to the other therapies which are entirely dependent upon individual interactions, a virtual community based treatment has been proposed in [Wu et al. 2010]. A virtual community consists of doctors, participants, and virtual space. Patients are often scared of real world, so this kind of community gives him a chance to interact with real world and show his unique talent. Different kinds of behavior and interaction between patients and virtual communities allow experts to analyze them differently and develop a healthcare plan according to the behavior shown.

A recent proliferation of dolphin-assisted autism therapy has become a focus area for public and researchers both. This therapy involves the interaction of autistic children with dolphins which may be on pool activities or out of pool activities. On pool activities include touching, patting, hand gesturing, and certain other tricks with the real dolphins; while out of pool activities include watching videos, listening recorded sound, touching and feeling the unreal dolphins which may be some model of dolphin prepared with the fibers. These set of out-of-pool activities prepare children to encounter the dolphins in reality. But it is reported that playing or swimming with real dolphins can potentially harm children and also it is not always available and affordable to everyone. The virtual reality concept is an emerging field that can remove the above limitations and studies Researchers [Lahiri et al. 2011][Parsons et al. 2004][Lahiri et al. 2013] have shown that virtual reality concept has been widely successful and will play a significant role in developing various learning and communication skills in autistic children [Cai et al. 2013].

## 6. ADULT/WOMEN HEALTHCARE

Women's health care is most important element of their lives shaping their ability to care for themselves and their families. According to the data presented in "Women's Health Care Chartbook-2011" [Ranji and Salganicoff 2011] while most of the women in US enjoy good health, more than one third of them almost 35% report chronic health conditions such as diabetes or hypertension and one fourth suffer from depression or anxiety. Poor women (33%) are at higher risk of health problems than high income women (11%) and also African women are more prone to diseases than white women. The rate of poor health, chronic diseases and disability increases as women get older which limits their daily activities.

There are pervasive healthcare solutions [Kanno et al. 2007][Oda et al. 2007] to calculate menstrual cycle and next ovulation period for women using wearable temperature sensors. Also, there are systems [L. Chen et al. 2011][Marques et al. 2000] for pregnancy time maternal and child health monitoring and care using different health sensors. More such systems have been summarized in

Table 8.

In the rest of this section, we shall discuss several widely experienced diseases common in both men and women of different ages.

### 6.1 Cardiovascular Disorders

Cardiovascular diseases (CVD), are a group of disorders of the heart and blood vessels. People in low income countries are more exposed to risk factors causing CVD, which results in to 80% of global deaths in low and middle income countries having equal proportion in both male and female. Multiple pervasive healthcare techniques to detect and prevent CVDs have been discussed in Table 9.

Among the traditional techniques to monitor HRs are Holter monitors and ECG machines. Common Holter-based portable devices continuously monitor various electrical activities of cardiovascular system for at least 24 hours. However, these recordings are not only quite tedious and time consuming to analyze but also they do not provide any real time feedback [Z. J. Z. Jin et al. 2009].

Electrocardiogram (ECG) is the most widely used clinical tool to measure the abnormal rhythms of the heart, and helps to diagnose properly. ECG machines are commonly used in hospitals and supervision is stationary by attaching electrodes to the body. But HR may not be abnormal while visiting a doctor [Z. Jin et al. 2009]. It may occur any time and ECG needs to be performed at the right moment to capture the snapshots in real-time. Also, ECG waveforms vary widely in nature, and in order to detect potential abnormalities, we need a system which can classify every possible ECG wave. However, that is a daunting task because no matter how much training data is used to train the classifiers it is hardly possible to cover every patient's ECG waveform. So some patient adaptable ECG classifier is required which is cost effective, portable and accurate. An effort towards that has been made [Hu et al. 1997] by proposing three ANN related algorithms, namely, self-organizing maps (SOM), learning vector quantization (LVQ), and mixture-of-experts (MOE) approach. They provide a personalized healthcare by developing an ECG classifier that further improves the accuracy of ECG processing. Several other ECG beat classifier algorithms [Osowski and Linh 2001][Dokur and Olmez 2001][Engin 2004][Güler and Übeyli 2005] have been developed using fuzzy hybrid neural network. However, making classifiers that work equally well on all datasets is still a problem.

Pervasive healthcare can monitor vital signs in real time and raise alarms when something abnormal is noted. This may save lives both by alerting people to risks they are exposed to and by arranging medical care as and when required. Several systems [Z. J. Z. Jin et al. 2009][Z. Jin et al. 2009][Chen et al. 2007][Ho and Chen 2009][Lee et al. 2011][Oliver and Flores-Mangas 2006] have been developed which use a set of vital sign sensors (ECG sensors, electrodes, oximeter, accelerometer) interconnected to form a BAN and they report their readings to the user's smartphone. The smartphone is often connected with a back-end server for big data analytics. Depending on the available resources, the smartphone can either send all the sensed data [Lee et al. 2011] to the back-end server or it can only send the abnormal data [Chen et al. 2007][Ho and Chen 2009] filtering the rest. Smartphone receives data from sensors via Bluetooth and forwards to the back-end server through GPRS or radio frequency. Many existing system [Klug et al. 2010] uses smartphone just for displaying the graphical results or the report in convenient manner.

Myocardial infarction (MI) (known as *heart attack*) occurs when normal blood flow stops in a part of the heart, due to narrowing or blockage of coronary arteries, causing necrosis (premature death of cells in living tissue) of heart muscles. Typical equipment for early diagnosis consists of Central Laboratory Testing (CLT) and Point-of-Care Testing (POCT). CLT provides accurate results incurring long testing delays, while POCT delivers quick but often inaccurate results. Research [Lee et al. 2011] has been carried out to develop a body-mounted POCT device which uses a micro-needle to collect blood sample and to detect biochemical markers related to heart attack within 15 minutes.

### 6.2 Cancer

Cancer is the uncontrolled growth and spread of cells which can affect almost any part of the body. Cancer accounts for 7.6 million deaths (around 13% of all deaths) in 2008 and more than 10 million new cases come up every year. Most cancer deaths are caused by lung, stomach, colon, liver, and breast cancers. Pervasive healthcare research on cancer detection mostly depends on several data mining and decision making techniques.

Several decision support systems (DSS) have been used successfully in detecting different kinds of cancers. An accuracy of 96% was achieved using decision-tree learning to analyze mass spectra of prostate cancer patients [Adam et al. 2002]. An accuracy between 83-100% was achieved in applying a three-layer perceptron ANN with a back propagation algorithm to analyze mass spectra for predicting astroglial tumor grade [Ball et al. 2002]. SVM is used for cancer detection [Li et al. 2004] using serum proteomic profiling. A genetic fuzzy system [Shabgahi and Abadeh 2011] has been developed to detect cancerous tumor by exploring gene expression data. Genetic algorithm and self-organizing cluster analysis was used to detect ovarian cancer [III et al. 2002] with a satisfying result. Automated detection system for cervical cancer using digital colposcope can be used by trained workers to the affected people of remote locations [Das et al. 2012]. Liver cancer detection system [Pinheiro and Kuo 2012] is developed using data mining techniques Apriori and FPGrowth.

Mammogram breast X-ray imaging is considered the most effective, low cost and reliable method in early detection of breast cancer. A system [Mohanty et al. 2012] has been developed for early breast cancer detection, using mammogram digital imaging and association rule classifier which gave very good resullt. There is also a computer-aided breast cancer detection system [Alolfe et al. 2008] which classifies normal and cancerous tissues in digital mammograms using the K-nearest neighbor (K-NN) technique and gives the best result with k=1. Radar-based breast cancer detection system [Klemm et al. 2009] uses a Hemispherical Antenna Array and the beam-forming technique to successfully detect 4 and 6 mm diameter spherical tumors in the curved breast phantom.

Authors of [Delen et al. 2005] have compared the performance of ANN, decision tree, and logistic regression based techniques of predicting breast cancer survivability where the decision tree performed the best and the logistic regression performed the worst of the three models evaluated.

### 6.3 Diabetes

Diabetes is a chronic disease that occurs when either the pancreas does not produce enough insulin, or our body cannot use the produced insulin effectively. Raised blood sugar is a common effect of uncontrolled diabetes and over time it seriously damages the nerves and the blood vessels and may even lead to blindness, amputation, and kidney failure.

Telematic expert system Diabeto [Turnin et al. 1992] is a tool for diet self-monitoring for diabetic patients which increases dietetic knowledge of patients and improves their dietary habits. Web-based system for decision support and tele-monitoring of diabetes patients help in their self-management and provide a way of dealing with the complicated health issues [Ahmed 2011][Roudsari et al. 2000].

Several systems using decision support techniques have been developed for the diagnosis and prediction of diabetes mellitus.

Case-based reasoning [Chen et al. 2010] stores past records and solves new problem based on them. This requires maintaining a huge database to provide full personalized recommendations. In order to overcome this drawback, ontology is used to reason about diabetes care concepts in case the case-based reasoning module cannot find satisfying cases. Sometimes, classical ontologies cannot sufficiently handle imprecise and vague knowledge for some real world applications. However, a system for fuzzy ontology can effectively resolve data and knowledge problems with uncertainty. A five layer ontology is developed in the fuzzy expert system to describe the knowledge with uncertainty for diagnosing diabetes [Lee and Wang 2011]. A neural network based approach [Singh and Kumari 2013] for automated detection of Diabetes Mellitus achieves an accuracy rate of 92.8%.

Association rule and Bayes network based techniques [Kasemthaweesab and Kurutach 2011][Hu et al. 2012] predict diabetes mellitus-2 with an accuracy of 72.3%. SVM is used in predicting diabetes [Barakat et al. 2010] with an accuracy of 94%.

E-NOSE detects the possibility of diabetes [Siyang et al. 2012] on the basis of urine odor. Artificial urine was simulated for a diabetes patient by adding various concentrations of glucose into the pure urine samples. 8 TGS 826 commercial chemical sensors (gas sensor array) were used as the sensing elements of E-NOSE.

### 6.4 Asthma

Asthma is a chronic disease characterized by recurrent attacks of breathlessness and wheezing in which the patients' lung capacity is significantly diminished. During an asthma attack the bronchial tubes swell narrowing the airways and reducing the flow of air into and out of the lungs. Recurrent asthma symptoms frequently results into sleeplessness, daytime fatigue, and reduced activity levels. It is also the most common chronic disease among children.

Systems have been developed using different technologies and methodologies to diagnose, monitor, and manage asthma.

MBreath [Al-Dowaihi et al. 2013] is an asthma monitoring system which allows patients to self-monitor their symptoms and act accordingly. It also enables patients to inform the medics in case of emergency by using an Android mobile application and a web portal for medical staff. Another portable wireless sensor based asthma monitoring system [Cao et al. 2012] measures user's respiratory airflow, blood oxygen saturation, and body posture using a respiration-posture sensor node and an oximeter sensor node.

A smartphone based asthma monitoring system developed especially for children is DexterNet [192] which has a hierarchical system architecture to monitor and manage asthma using a Nokia 810 smartphone and an Aerocet 531 particle counter (to monitor air pollution exposure). The smartphone acts as a gateway of a Bluetooth connected BAN which has GPS, accelerometer, and gyroscope to monitor the activity, location, and air pollution exposure of the child. At the highest level many patient's data is shared through a global network which also connects care givers and asthma researchers.

MobileSpiro [Gupta et al. 2011] is a smartphone based self-testing and self-management system for asthma patients. Spirometer is a device to measure the lung capacity of an asthma patient and can help check impending asthma attacks. Due to the difficulty in self-monitoring spirometer readings, many such tests fail to accurately capture lung capacity. In order to do away with the erroneous maneuvers by patients, MobileSpiro introduces an automated algorithm which works with more than 95% accuracy.

Another smartphone based asthma warning system [Chu et al. 2006] tracks user's outdoor location and warns about possible polluted locations by consulting with a remote server which has hourly air-pollution records of different locations.

An asthma monitoring system using audio data collects cough samples of asthmatic and non-asthmatic children using microphone. Then the features of cough sounds were extracted with the help of MATLAB which helps in diagnosis of asthma. The proposed system was tested on 20 samples which correctly classified 85-90% of the records.

Exhaled nitric oxide (NO) can also be used to detect asthma [Shelat et al. 2012] with the help of *electronic nose* (E-Nose) which is an array of chemical sensors. The NO level in asthmatic subjects were found to be higher than normal subjects.

### 6.5 Sleep apnea

Sleep apnea is a chronic condition characterized by infrequent breathing during sleep. Each pause in breathing may last from few seconds to minutes and may occur 1-100 times in an hour. With no or very little air passing through the blockage to the lungs, it may cause snoring and lower the blood oxygen level which usually disrupts the sleep. Usually with a loud snort or choking sound, the blocked airway opens and normal breathing resumes.

Most common type of apnea is the obstructive sleep apnea and it increases the risk of cardiac arrest, stroke, obesity, diabetes, headache, and hypertension. [Brooks et al. 1997][Mannarino et al. 2012]. This disorder generally goes undiagnosed because it canot be detected during visits to the doctor or by blood tests.

Polysomnography [Flemons and Mcnicholas 1997] is a standard test for sleep apnea which requires the user to sleep in the hospital wearing many sensing devices. This test is very expensive as it requires sophisticated instruments and trained clinicians. A less obtrusive and non-invasive monitoring system is Actigraph [Sadeh and Acebo 2002] which is to be worn on the wrist and it records the sleep-wake patterns of the user.

Detecting sleep apnea using pervasive healthcare techniques consist mainly of three approaches. The first one is detecting and analyzing body movements and posture during sleep. The second one is measuring sleep-time blood oxygen level. The third one is measuring cardiac interbeat interval (IBI) during sleep. Both the approaches require using appropriate wireless sensors and body area network.

Based on the first approach, a sleep monitoring system [Hoque et al. 2010] has been developed based on Wireless Identification and Sensing Platform (WISP). WISP tags are attached to bed mattress and accelerometer data is collected to analyze the body movements during sleep. The system uses three tags to record the body movements and tries to classify sleep-time body postures over bed using different categories, like empty bed, lying over bed, lying on back, lying on left or right side, etc. More accurate results can be obtained by using more number of tags and by placing them in proper positions.

Using the second approach, pulse oximetry is used for determining the saturation of oxygen in blood [Oliver and Flores-Mangas 2007]. Usually oximeter's measurements are based on the physical phenomenon that oxygen level present is determined by the amount of light absorbed in hemoglobin.

Cardiac interbeat interval time series measured using ECG is used as the third approach to detect and diagnose obstructive sleep apnea [Mietus et al. 2000]. This is because, the HR of sleep apnea patients shows cyclic increases and decreases followed by the resumption of breathing. Another similar approach [Delibasoglu et al. 2011] prepares the time series of instant HRs (IHR) derived from overnight sleep ECGs which are then analyzed using wavelet decomposition. This approach tries to find a reliable and practical technique to detect sleep apnea by measuring the real time occurrences of sleep-disordered-breathing (SDB). This approach is implemented on segments of 6-minutes length IHR in which $4^{th}$ minute is considered apnea deciding minute. Wavelet decomposition has a success rate of over 96.6% in deciding whether a person have apnea or not.

HealthGear [Oliver and Flores-Mangas 2007] is a wearable and real time sleep apnea monitoring system which can detect sleep disorder like, snoring. However, HealthGear is just for detecting a disorder and some other technique is required to diagnose the problem.

A smartphone based approach to aimed at curing sleep apnea is the auto adjustable pillow system [Zhang et al. 2013]. This uses three blood oxygen sensors mounted on the user's fingertip to keep monitoring sleeping disorder by measuring blood oxygen level and pulse during night. The smartphone is used as a central controller which controls the adjustment of pillow on the basis of real-time feedback from a pillow adjustment algorithm. The system is divided into 4 modules; signal receiver, apnea detector, pillow controller, and presentation module. The last one can show the real time information, like oxygen level, apnea, and pillow status to the user and his family members.

We have provided a research summary of pervasive healthcare systems addressing sleep apnea in Table 10.

### 6.6 Stress

Stress is a physiological response to the mental, emotional, or physical challenges that we all encounter at some point of time. During stress, our body secretes hormones, such as adrenaline, into the bloodstream which increases HR and BP and results in altered immune system.

Acute (short-term) stress leads to rapid changes throughout the body which may develop many physical and psychological problems. Chronic (long-term) stress can have real health consequences, like hypertension or coronary artery diseases [Kubzansky and Kawachi 2000][Holmes et al. 2006][Pickering 2001] .

One of the conventional methods to characterize stress is conducting interview during a visit to an expert. But this approach gives a momentary snapshot which may not be same as actual patterns followed by daily mental or physical activities of the user. That is why continuous monitoring is essential for understanding and managing personal stress levels [Sun et al. 2012]. Physiological sensors can monitor user continuously and transfer the data to the smartphone carried by the user.

Smartphones are very useful [Bauer and Lukowicz 2012] for monitoring stress related changes in human behavior through different embedded sensor nodes. GPS and Wi-Fi are used to track user locations which are usually different for a user at stressed and stress-free times. Bluetooth is used to track user's social interactions which are usually more when the user is stress-free. Call behavior (whom the user contacted, how frequently, and what is the relationship) and SMS frequency are used to monitor user's behavior. Experiment on some students monitored during pre- and post-exam periods shows that the change in the behavioral pattern varies from person to person. StressSense [Lu et al. 2012] is another smartphone based system for stress detection which uses the Google nexus android smartphone and its microphone and video camera to detect stress. StressSense achieves an indoor and outdoor accuracy of 81% and 76%, respectively using GMM classification framework.

Physiological sensors such as Galvanic skin response (GSR), ECG sensors, respiration sensors, and skin temperature (ST) sensors are used in monitoring mental stress [Sun et al. 2012][Zhai and Barreto 2006][Choi and Gutierrez-Osuna 2009][de Santos et al. 2011][A. D. S. Sierra et al. 2011][A. de S. Sierra et al. 2011][Singh et al. 2011][Mokhayeri and Akbarzadeh-T 2011]. The advantage of physiological sensors is that they are unobstructive, non-intrusive and non-invasive. However, they require direct interaction between human and sensors. The data collected by physiological sensors are analyzed using several data mining techniques, such as SVM, J48 decision tree, Bayes network, and fuzzy logic. Real time stress-level determination in drivers has been experimented in [Healey and Picard 2005] using ECG, EMG, GSR, and respiration sensors with an accuracy as high as 97%.

However, physiological signals like HR and skin conductance solely may not be suitable for stress detection as the HR may also increases due to physical activity. Segregating stress and exercise related changes in vital signs has been researched in [Sun et al. 2012] and the authors propose to use an accelerometer to monitor physical activities along with the ECG and GSR sensors which monitor the stress level. Collected data is classified using decision trees, Bayesian network, and SVM. Experiments conducted on 20 subjects, across 3 activities, sitting, standing, and walking, shows that by removing accelerometer, accuracy of stress detection in mobile environment decreases significantly. Decision tree classifier gives highest accuracy of 92.4% combining all features.

Another standard approach for detecting stress works by measuring changes in certain vital signs under duress/mental load using different sensor nodes [Zhai and Barreto 2006][Choi and Gutierrez-Osuna 2009]. Then the collected data is classified using several data mining techniques into normal and stress conditions. One way of creating a stressful situation is through the classical stroop test carried out in an interactive manner, in which a user clicks on correct answer rather than stating it verbally. [Zhai and Barreto 2006] measures user's vital signs using GSR, Blood Volume Pressure (BVP), Pupil Diameter (PD), and ST sensors during a stroop test and then apply SVM to classify the state as *stress* or *normal*. Results show that the PD is the most significant feature in correctly determining mental stress. Another similar approach [Choi and Gutierrez-Osuna 2009] works by measuring HRs of user's while taking stroop Color Word Test (CWT) or Mental Arithmetic Test (MAT) and two other tests to induce relaxation, like deep breathing. The stress detecting system identifies stressful events with a high accuracy. One more approach [Mokhayeri and Akbarzadeh-T 2011] uses stroop CWT to produce stress situation. It utilizes video camera to monitor eye movements and ECG and PPG sensors to monitor physiological signals. After removing noises, like eye blinks using fuzzy filter, normalized pupil images are obtained. Pupil diameter and dilation accelerations are also calculated. Finally, fuzzy SVM is used to classify the features into stressed and non-stressed categories. This research also identifies the pupil parameters as the most important feature for stress recognition.

Different pervasive healthcare techniques to detect stress have been discussed in Table 11.

### 6.7 Obesity

Obesity is regarded as a global epidemic resulting in many abnormal health conditions including heart disease, diabetes, various cancers, liver and gallbladder disease, sleep apnea, respiratory disease, reproductive health complications, and mental health disorders. Intake of high calorie food and beverages and sedentary lifestyle are the leading causes of obesity and related abnormalities [Khan and Siddiqi 2012]. Experiment performed on 8 severe obese (BMI >= 40.0kg/m$^2$) candidates have shown that after losing weight, their HR lowered significantly which clinically indicates decrease in heart risk [Poirier et al. 2003].

There are two ways to be fit; either user needs to be self-motivated for his fitness or he should be monitored continuously, so that, calorie intake approximates the calories expenditure. Pervasive healthcare research on tackling obesity has focused on these two pronged approaches, the first is *diet monitoring* and the second is *urging the user to carry out fitness activities*.

Diet monitoring is usually done by actually identifying food/drink intakes through camera or by indirectly detecting the eating gestures. Nutritional information of food items need to be captured from the information provided on the containers through bar code scanning or other viable methods. Gesture or activity recognition systems using wearable sensors, like accelerometer and gyroscope can be used in automatic diet monitoring [Amft et al. 2005][Junker et al. 2004]. There are also many other systems [Khan and Siddiqi 2012][Sha et al. 2008][Denning et al. 2009][Consolvo et al. 2008] using accelerometer or gyroscopes which sense the ADLs, like sitting, standing, walking, running, etc. and calculate the energy expenditure based on some pre-defined standards [Bourke et al. 2007]. The system described in [Amft et al. 2005] uses motion sensors attached to wrist and upper arm to differentiate eating and drinking gestures from other arbitrary gestures using HMM with an average accuracy of 94%.

In BALANCE [Denning et al. 2009] the calorie content of a food item has to be manually entered, but the system is capable of automatically detecting the calorie spent in daily activities with the help of inertial sensors worn on the body. There are smart systems capable of collecting nutritional information of food items through barcode scanning or voice recording. A PDA based application [Siek et al. 2006] has been developed to assist people with low literacy skills and suffering from chronic kidney diseases to monitor their food intake and their nutrition values. HyperFit [Järvinen et al. 2008] is a system which uses smart phone camera attached with a special lens for barcode scanning of food items. Both systems discussed above only allow diet monitoring while BALANCE [Denning et al. 2009] also takes into account the kind of activities the user is performing.

There are several smartphone based systems developed for diet monitoring that use pictures taken using a smartphone camera which capture the quantity of food intake [Järvinen et al. 2008]. DietSense [Reddy et al. 2007] uses ImageScape software to process large set of such images. There is a PDA based system [Wang et al. 2006] which uses camera to take the pictures of daily food items and send it to the dietitian for further assessment. HealthAware [Gao et al. 2009] uses accelerometer and GPS embedded in a smartphone to monitor activities and the camera to analyze food items intake. User needs to manually enter name of a food item and the system will calculate the calorie based on stored database.

The second approach to control obesity works by encouraging the user to carry out fitness activities. TripleBeat [Oliveira and Oliver 2008] uses real-time musical feedback and persuasion technique to encourage a user to achieve his fitness goal. The system consists of ECG sensors to monitor the HR and ability to pace and accelerometer to measure movements during run. Another such system is UbiFit Garden [Consolvo et al. 2008] which uses smartphone to encourage healthy life style. UbiFit Garden uses classifiers that are trained to differentiate walking, running, cycling using an elliptical trainer and stair machine with the help of 3-d accelerometer and barometer. ExerTrek [Ho and Chen 2009] is a system for monitoring exercise and exercise-time HR and it provides real-time online feedback immediately to the user regarding their heart status and any abnormalities.

There are many commercial mobile apps available in the market which tracks user activities and helps them to shed extra weights. Nike+ Running, RunKeeper, Sports Tracker, Sportypal are some of the free apps available for apple iPhone that can record distance, pace, and duration of running while some of them provides audio feedback at every mile. Some of these apps provide the facility to tag the Facebook friends and to share a map of route travelled during exercise with friends and family. Runtastic PRO and Endomondo Sports Tracker PRO are some of the paid fitness monitoring apps available for iPhone and Android.

A brief summary of pervasive healthcare solutions to monitor and control obesity have been discussed in Table 12.

## 7. DISABLED HEALTHCARE

The International Classification of Functioning, Disability and Health (ICF) define disability as "an umbrella term for impairments, activity limitations and participation restrictions". Disability can be congenital or it can be introduced due to some illness or falls at a later stage of life. Now a days, the rates of disability are increasing in part due to ageing populations and an increase in chronic health conditions. People with disabilities have less access to health care services and therefore experience unmet health care needs. Pervasive healthcare can provide a continuous monitoring and support service for disabled individuals. An extensive coverage of multiple disabled healthcare systems have been provided in Table 13.

Bus identification system [El Alamy et al. 2012] for visually impaired person uses radio frequency tags and reader to help visually impaired patients. Another system called PERCEPT also uses RFID technology for indoor navigation system for visually impaired or blind persons [Ganz et al. 2012]. The system Drishti [Hela et al. 2001][Ran et al. 2004] is for both indoor and outdoor navigation of a visually challenged person. Ultrasound tags and GPS are used as sensors in this system. Another system [Roberts et al. 1999] for disabled people suffering from spina bifida uses infrared remote control to reduce the pain to some extent.

## 8. OPEN CHALLENGES

Pervasive healthcare systems face many challenges. They are partially addressed [Al Ameen et al. 2012][Meingast et al. 2006][Liang et al. 2012][Alemdar and Ersoy 2010], but still a lot of them remain to be addressed. We elaborate them in the remaining sub-sections.

### 9.1 Security and Privacy

Patient's vital signs and medical records captured by health sensors are transferred via different wireless connections and through Internet. This involves plenty of security and privacy risks for the patients. An attacker can eavesdrop on wireless channel to listen or to modify user's private health data with malicious intent. The security requirements of healthcare systems are confidentiality, data integrity, accountability, availability, and access control. Proper user authentication and authorizations should be included to ensure better control of rightful user over his private data. Moreover, the WBANs consisting of multiple third party health sensors make it easy for malicious adversaries to launch security attacks which include compromising or paralyzing body sensors and WBAN gateways. Also, applications which track the location of a patient or person if compromised may lead to grave consequences. So, challenges lie in protecting the privacy and confidentiality of sensitive health information of individual user's with efficient encryption techniques.

Also, to ensure security over the medley of third party wireless sensors is extremely challenging.

### 9.2 Reliability

Reliability should be ensured at different stages, such as, data acquisition, data transmission, and data processing. Sensors responsible for data acquisition are resource-constrained and failure-prone devices. At times, they behave arbitrarily and may produce incorrect readings. Also, wireless communication channels are error-prone due to interference and path fading, and multi-hop routing paths are not stable because of user mobility. Due to all these reasons BANs may produce and transmit inaccurate information which can be misleading, resulting in inaccurate diagnosis, unnecessary monetary cost, and even loss of lives. Sometimes, multiple sensors are necessary to ensure data reliability not achievable by using single sensor node.

### 9.3 Efficiency / Power Management

Major components of a health monitoring system, such as the sensor nodes, wireless gateways, and the smartphones are resource-constrained entities with limited power supplies. The algorithms embedded in them must be highly efficient so as to enable long-term as well as outdoor monitoring and continuous data collection.

### 9.4 Mobility

Supporting user mobility is another important challenge for wireless health monitoring systems. Quality of life of users can only be maintained if the healthcare system does not restrict his mobility in the name of patient monitoring. Wireless BAN consisting of sensor and wireless technology mostly ensures user mobility. However, still many challenges remain in developing a fully-functional mobile wireless BAN with high

### 9.5 Heterogeneity

Ease of deployment is another challenging issue in sensor based healthcare system. Multiple third party sensor nodes with varied features are brought in the same platform to work together. New or replacement sensors brought in should be able to easily integrate with the existing infrastructure.

### 9.6 Miscellaneous

Health monitoring systems are meant for elderly individuals, children, or other patients and should be user-friendly and easy-to-use. They should be context-aware or context-triggered in which they start operating or fully functional on detection of abnormalities in particular vital signs. Even if the user forgets or becomes unable to provide necessary inputs, the system should have proper automation to act correctly on its behalf. Wearable sensors should be unobtrusive and largely non-invasive. Making pervasive healthcare solutions to each and every individual in need is another social challenge. Moreover, in rural and remote areas, poor or unavailable data communication services, like 3G may hamper data transmission to central servers.

## 9. CONCLUSION

In this paper, we have presented a comprehensive survey of different tools and techniques used in Pervasive healthcare in a disease-specific manner. We have covered the major diseases and disorders that can be quickly detected and treated with the use of technology, such as fatal and non-fatal falls, Parkinson's disease, cardio-vascular disorders, stress, etc. We have discussed different pervasive healthcare techniques available to address those diseases and many other permanent handicaps, like blindness, motor disabilities, paralysis, etc. Moreover, a plethora of commercially available pervasive healthcare products have been listed in Table 14. Our paper provides an overall understanding of the various aspects of pervasive healthcare with respect to different diseases. We have also identified several open challenges associated with pervasive healthcare research for the benefit of future researchers.

**Table 1: List of Acronyms**

| ACRONYM | FULL FORM | ACRONYM | FULL FORM |
|---|---|---|---|
| ADL | Activities of Daily Living | HR | Heart Rate |
| ALS | Amyotrophic Lateral Scelerosis | IR | Infra-red |
| ANN | Artificial Neural Network | K-NN | K-Nearest Neighbor |
| ASD | Autism Spectrum Disorder | LAN | Local Area Network |
| BBT | Basal Body Temperature | LHMM | Layered Hmm |
| BP | Blood Pressure | MEMS | Micro-Electro-Mechanical Systems |
| BSN | Body Sensor Network | PAN | Personal Area Network |
| BVP | Blood Volume Pressure (Sensor) | PCA | Principal Component Analysis |
| CDC | Centers For Disease Control And Prevention | PD | Parkinson's Disease |
| CHMM | Coupled Hmm | PD | Pupil Diameter (Sensor) |
| CVD | Cardiovascular Diseases | PPG | Photoplethysmogram |
| DBS | Deep Brain Simulation | RF | Radio Frequency |
| DWT | Discrete Wavelet Transform | RFID | Radio Frequency Identification |
| ECG | Electrocardiogram | RSD | Reflex Sympathetic dystrophy syndrome |
| EDA | Electro Dermal Activity | ST | Skin Temperature (Sensor) |
| EMG | Electromyogram | SVM | Support Vector Machine |
| EOG | Electro-Oculography | TAA | Tri-axial accelerometer |
| FDS | Fall Detection System(s) | UPDRS | Unified Parkinson’s Disease Rating Scale |
| GIS | Geographic Information System | WBAN | Wireless Body Area Network |
| GMM | Gaussian Mixture Model | WHO | World Health Organization |
| GPS | Global Positioning System | Wi-Fi | Wireless Fidelity |
| GPS | Global Positioning System | WISP | Wireless Identification and Sensing Platform |
| GSR | Galvanic Skin Response (Sensor) | WPAN | Wireless personal area network |
| HMM | Hidden Markov Model | WSN | Wireless Sensor Network |

**Table 2: List of Sensors Used in Pervasive Healthcare Applications**

| Sensors | Functions |
|---|---|
| Accelerometer | Measures linear acceleration of movement in 3-D space. |
| Gyroscope | Measures the orientation based on angular momentum. |
| Image Sensor | Converts an optical image into an electronic signal. |
| GPS | Space-based satellite navigation system that provides location and time information. |
| Ultrasonic Sensor | Calculates the distance to an object based on high frequency sound waves |
| Microphone | Converts sound into an electrical signal |
| Temperature Sensor | Measures temperature |
| Piezoelectric / pressure Sensor | Uses the piezoelectric effect to measure pressure, acceleration, strain or force |
| Ambient light Sensor | Used to detect light in a similar way as human eye. |
| Skin Electrodes | Detects electrical activity of the skeletal muscles |
| Arm Cuff-based Monitor | Measures pressure exerted by circulating blood upon the walls of blood vessels. |
| IR / Proximity Sensor | Detects the presence of nearby objects using IR spectrum |
| Magnetometer / Compass | Measures the strength and the direction of magnetic fields |
| Pulse Oximeter | Monitor patient's oxygen saturation in blood |
| Tactile Sensor | Sensitive to touch, force, or pressure |

**Table 3: Sensors Used and Problem Addressed**

| Paper | Sensor | Placement | Problem Addressed |
|---|---|---|---|
| [L.J.G. et al. 2011][Chuah and DiBlasio 2012][Amft et al. 2005][Noury et al. 2000][Bourke et al. 2007][Sposaro and Tyson 2009] [Lai et al. 2010][Jantaraprim et al. 2010][Dai et al. 2010][Zhao et al. 2012][P.N. et al. 2012][Smith and Bagley 2010] [Karantonis et al. 2006][Bourke et al. 2008][Caporusso et al. 2009][Fang et al. 2012][Selvabala and Ganesh 2012][Putchana et al. 2012] [Abbate et al. 2012][LeMoyne et al. 2009][Hu et al. 2013][Mariani et al. 2013][LeMoyne et al. 2010] | Accelero-meter | Head, neck, chest, waist, wrist, thigh, hands, shoes, Pants' pocket (Smartphone) | Computer interface for disabled using voluntary eye movement, Autism, Automatic diet monitoring, Indoor and outdoor fall detection, Monitoring of child with Cerebral palsy, PD |
| [Hela et al. 2001][Ran et al. 2004][Chuah and DiBlasio 2012][Hsiao et al. 2011][Lin et al. 2012] | GPS | Smartphone | Wireless indoor and outdoor navigation system for visually impaired and disabled, Autism, Outdoor fall detection, support elderly with dementia travelling outdoor |
| [Takami et al. 1996][Lee and Lee 2009][Subbu and Gnanaraj 2012][Tzeng et al. 2010][Song et al. 1998][Alwan et al. 2006] [Okuno et al. 2007] | Piezoelectric sensor | On glasses, Belt, Placed to form Braille display, Floor, Mounted on the end-effector of the robotic arm of KARES, Finger | for disabled people - interactive interface using head movement, for visually impaired - Indoor positioning system for moving objects for indoor sports, Electronic device to read messages from modern gadgets, Floor vibration based fall detector, PD |
| [Stickney et al. 1999][Lv et al. 2008][Yu et al. 2013] | Ultrasonic Sensor | 4 corners of child vehicle, Robot | Remotely operated toy vehicle for disabled child, Speech controlled robot for detecting obstacles in all directions, Safe walking for visually impaired |
| [Ishimatsu et al. 1997][Lim and Singh 2005][Su et al. 2008][Krejcar 2011][Anderson et al. 2006][Schulze et al. 2009] [Hsiao et al. 2011][Smith and Bagley 2010][Tanaka et al. | Optical Sensor/ Video sensor/ Image sensor | On Spectacles or hair band, Video Camera, Head and Mouth | Computer interface/ virtual control recognizing head/ eyelid/ vision/ slight body movement for disabled people/ patients of ALS and cerebral infarction, Activity recognition and fall detection, Monitoring of child with Cerebral palsy |
| 2008] | | | |
| [Tartamella et al. 1999][Khan and Enderle 2000] | Temperature sensor | Patient's wheelchair, | Remote control digital thermostat and door opener for disabled |
| [Carvalho et al. 1999][Liu et al. 2005] | Voice-sensor (microphone) | fitted to a standard room thermostat, Microphone connected to computer | Voice based control of room temperature / computers / robots for disabled, Multiple Sclerosis, and paraplegia patients |
| [Khan and Enderle 2000][Lee and Lee 2009][El Alamy et al. 2012] [Ganz et al. 2012][Noury et al. 2000][Chen and Wang 2007] | RF Technology | Remote attached to wheel chair, RF transmitters on head, RFID tag added in PDA, RFID reader in bus station, RFID tags in wrist | Reflex Sympathetic dystrophy syndrome (RSD), for visually impaired- Indoor positioning system for moving objects for indoor sports, Bus Identification, fall detection of elderly |
| [Tzeng et al. 2010][Selvabala and Ganesh 2012] | IR Sensor | Floor | Fall detection |

**Table 4: Communication Technologies Used in Pervasive Healthcare Applications**

| Communication Technology | Applications | Range | Features |
|---|---|---|---|
| IEEE 802.15.6 [Ullah et al. 2013] | WBAN or BSN | near and around the human body | Communication standard for low power devices implanted inside human body. |
| Bluetooth (IEEE 802.15.1) [Bisdikian 2001] | WPANs | 0-100 meters | Low cost, user friendly replacement for the interconnect cables between personal devices. Bluetooth network joining time is of the order of 3 sec. |
| Wi-Fi (IEEE 802.11)[Kuran and Tugcu 2007] | WLANs | 50-60 meters using 802.11g or 802.11n | High power consumption, Less secure than wired connection but high data rate (54 Mbps in 802.11g and 540Mbps in 802.11n) |
| ANT [Wetherall et al. 1998][Jayashree et al. n.d.] | WSN, WPANs, and WLANs | 30 meter line of sight | Low overhead, ultra-low power, interference-free characteristics, low data rate. Applications in the health, home automation, and industrial sectors. |
| RFID [Weinstein 2005][Garfinkel et al. 2005][Nath et al. 2006] | Readers identity electronic tags from a long distance | 20-100 meters for active RFID tags. No line-of-sight is needed. | Active tags are larger and more expensive, passive tags are smaller and inexpensive. Tags can be read by other than owner without any knowledge. |
| Infrared | Bi-directional data transfer for computing devices | Up-to 5meters only in the "line of sight". IR signals cannot penetrate obstructions. | Supports up-to 4 Mbps data rate. |
| ZigBee (IEEE 802.15.4) [Salman et al. 2010] | WPANs | 10-100 meters | Low powered, decentralized architecture, secure networki-ng using 128 bit symmetric encryption key. More battery life than Bluetooth devices. Network joining time ~30 msec |

**Table 5: Fall Detection Research Summary**

| Author | Sensor | Placement | Problem addressed | Advantages | Disadvantages |
|---|---|---|---|---|---|
| [Noury et al. 2000] | Wireless position sensor , Fall sensor Actimeter | Each Room has a sensor | FD*, Remote monitoring of patients | Provide useful lifestyle and behavior data to the physician in charge of the remote care of the patient. | Setups are limited to only one room |
| [Alwan et al. 2006] | Piezoelectric sensor | Kept on floor | Floor vibration based FD | Passive and un-obstructive, Can distinguish human and object falls | Detection distance up to 20 feet, need multiple detector for each room |
| [Karantonis et al. 2006] | TAA | Waist mounted | Acceleration based FD | Can distinguish between activity and rest and recognize sitting, standing, and lying postures with high accuracy | Need to mount on waist every time |
| [Anderson et al. 2006] | Video sensor (webcam) | Room | Image based (Silhouette) FD | The privacy of residents is ensured through the extraction of silhouettes | Limited area covered |
| [Williams et al. 2007] | Video sensor (video camera) | Room | Image based FD | Low power cameras are used so low power is consumed | 5-6 cameras needed per room, Privacy concern |
| [Lin and Ling 2007] | Video sensor (video camera) | Room | Compressed-domain vision-based FD | Can differentiate fall-down from squatting considering event duration | Require prior setup and maintenance of system |
| [Bourke et al. 2007] | TAA | Trunk or thigh | Threshold based FD | Simple, works well for resource-constrained wearable devices | Cannot send emergency message. Always need to wear with some clothes |
| [Chen and Wang 2007] | Wrist-tag with 2 way radio, active RFID readers | wrist | RFID based FD | Can call for real time help by pushing the wrist-tag, 2 way communication | Pushing wrist tag button may not be feasible in critical situations |
| [Bourke et al. 2008] | TAA | Vest under clothing | FD | Vest is a light-weight (320g) garment made from 100% polyester mesh. | Wearing the vest every time is uncomfortable. |
| [Caporusso et al. 2009] | TAA | wrist | Risk awareness for fall prevention | Wearable and non-intrusive, system learns quickly from the patients | Wearing it always on the wrist is not feasible. |
| [Schulze et al. 2009] | Video sensor (video camera) | Room | Image based FD | The camera is ceiling mounted and completely covers the room | Human shadow produced in bright daylight hamper tracking and lead to undetected falls or false alarms. |
| [Sposaro and Tyson 2009] | TAA, GPS, smartphone | Pants' Front pocket | Threshold based FD | Less intrusive device, less false positives due to voice comm. between the fallen-person and his social contact | Higher level of activity if the phone is carried on more accelerated body parts, like the arms |
| [Shieh and Huang 2009] | Video sensor (video camera) | Room | Image (human shape) based FD | Apply pipelining and multithreading technique to improve throughput of processing multiple video streams | Assumption for true fall: person cannot stand up in 10 seconds. It might create false alarms |
| [Smith and Bagley 2010] | TAA and video sensor | Child wear it daily in a small fanny pack | Assess the impact of treatment on ADLs in children with Cerebral palsy using threshold based FD and activity recognition system | Pre-therapy and post-therapy recordings from monitors will be used to assess the efficacies of alternative treatments for Gait abnormalities | Child may not always wear it and so accurate data will not be recorded. |
| [Lai et al. 2010] | accelerometer, | Neck, waist Left/right-wrist, left/ right-thigh | Multiple acceleration based FD | Self-learning system - so detection rules can be adjusted for each person's specific motion behaviors | Cannot correctly determine the situation of body after collision, person-specific - cannot be done using single sensor system |
| [Diraco et al. 2010] | Video sensor (video camera) | Room | Active vision system for FD and posture recognition | Guarantee privacy, camera can sense multiple events simultaneously, videos are used for post-fall analysis | Video sensing is suitable for indoors only |
| [Tzeng et al. 2010] | Pressure sensor, IR sensor | Room floor | Floor pressure and Threshold based FD | User need not wear any sensor | Infrared cameras form a thermal image of human body which may be hampered by wearing clothes |
| [Abu-Faraj et al. 2010] | Video sensor, capacitive sensor | Room | Fall detection and recovery in young able adults using dynamic planter pressure measurement | The falls in young adults could be prevented via exercise intervention programs, including resistance, endurance, balance training, and other therapies | Uses previous set-up to find the fall recovery time against violent tremors which may be far different than natural fall and fall recovery in tremors |
| [Jantaraprim et al. 2010] | TAA | trunk | Improves FD accuracy using 2 thresholds - *free fall* and *beginning to max peak*) | 2 threshold FD algorithm has low computational complexity, allowing its real-time implementation on a microcontroller | Data is processed offline |
| [Dai et al. 2010] | Accelerometer, Android G1 phone | Shirt/pants pocket, waist | Pervasive FD | Few false positives and false negatives. Available both indoor and outdoor. No extra hardware and service cost. Lightweight and efficient. | There is no way to distinguish between real fall and the phone falling accidently |
| [Hsiao et al. 2011] | ECG sensors, GPS, video sensor (video camera) | A healthcare box per patient | Outdoor FD system | Using GPS patients' location can be found and ECG will help to know patents' situation | Users always have to carry an external health-care box. |
| [Abbate et al. 2012] | TAA, Shimmer2 sensor, Nexus One smartphone | waist | FD and movement in elderly | Possibility of using a small external sensing unit can greatly reduce the intrusiveness of the system. External sensor improves the battery lifetime. | Women do not consider comfortable wearing a device on their belts, as they seldom use belts. |
| [Fang et al. 2012] | TAA, AndSP** | Chest, waist, thigh | FD and movement in elderly | Accurate detection | Require previous camera or environmental set-up and maintenance of whole architecture |
| [Zhao et al. 2012] | TAA and Wi-Fi | waist | Smartphone based FD | Using Wi-Fi gives better performance for indoors than systems using GPS | If user falls in new location where Wi-Fi access point is not in radio map then it cannot find user quickly |
| [Selvabala and Ganesh 2012] | TAA and Passive IR sensor | Environ-ment | WSN based human FD system | Do not affect privacy as sensors are fit into environment | Lot of sensors are needed to fit into environment to track user activities everywhere |
| [P.N. et al. 2012] | TAA and orientation sensor, AndSP | Chest or pants pocket | Semi-supervised FD algorithm based on features of fall events | The semi-supervised algorithm requires less training every time we want to deploy in a smartphone for different users | Training process can be exhaustive and resources consuming. |
| [Putchana et al. 2012] | TAA | Belt | FD and movement in elderly | Using ZigBee communication makes the system more efficient and less costly | younger subjects contribute data mimicking movements of older persons – this may introduce errors |

* Fall Detection / prevention – FD / FP, ** Android Smartphone - AndSP

**Table 6: Parkinson's Disease Research Summary**

| Author | Sensor | Placement | Problem addressed | Advantages | Disadvantages |
|---|---|---|---|---|---|
| [Wilson and Atkeson | 24 motion detectors and 24 contact switches, | Room, doorway, refrigerator, kitchen cabinets, | Simultaneous tracking and activity recognition (STAR) of elderly in room | Many people are uncomfortable living with cameras and microphones so non-invasive | Tracks only at room-level. Activity recognition is limited to whether or not an occupant is |
| 2005] | unique | drawers etc. | | sensors are used | moving |
| [Okuno et al. 2007] | TAA, touch (3-axis piezoelectric) sensor | 3 sensors worn on the index finger and thumb on finger stalls | Measurement system of finger-tapping contact force for quantitative diagnosis of PD | 3 low weight sensors are worn on fingers which is easy to use than some other systems | Wires attached to sensors for the transfer of data to PC may be uncomfortable to the user and is not easily portable |
| [Ahamed et al. 2007] | Sensors for Diabetes, obesity, irregular heart-beat, hypertension | Sensors attached to body to form body network | Wellness assistant integrating sensing, communication, and event management for healthcare of elderly people | Adaptable to multiple situation requiring periodic monitoring in different diseases and provide scheduled report | Lot of sensors will be required to monitor user which may be uncomfortable to wear |
| [Shima et al. 2008] | Magnetic sensors (MS) | 2 MS attached to the distal parts of thumb and index finger | Quantitative measurement and evaluation of finger tapping movements for diagnosis support and assessment of PD | Easier to use as only finger tapping is needed to check the symptoms of PD | Attaching magnetic coils and transferring the output voltage to PC don't give the freedom to use system anywhere, anytime |
| [Shima et al. 2009] | magnetic sensor, a three-axis force sensor | Force sensor is pinched against finger pads | Estimation of Human Finger Tapping Forces between two fingerpads using a Fingerpad-Stiffness Model without attaching a sensor to the finger | Finger tapping is very easy method as can be performed at any time at any place | A special device consisting of magnetic sensor, force sensor and multi-telemeter need to set-up |
| [Cunningham et al.2009] | ---------------- | ------------ | Identifying fine movement difficulties in PD using a computer assessment tool | Easy to use as just one PC with specific software is needed to assess the severity of PD | Non-computer literate people feel uncomfortable to click and that obviously affects the result |
| [LeMoyne et al. 2009] | 3D MEMS accelerometer | dorsum of hand | Quantification of PD character-istics using wireless accelero-meters in their natural home-based autonomous environment | Accelerometer is attached to hand which can be used very easily without creating a special set-up for this purpose | accelerometer is attaching to hand with a strap not integrated into glove |
| [LeMoyne et al. 2010] | iPhone 3-D accelerometer | Dorsum of the hand | Characterize PD tremors through a wireless accelero-meter application | iPhone 3G has 8 GB of storage so in case of no network data can be stored locally | An extra glove is needed to mount the phone on hand |
| [Kupryjanow et al. 2010] | tactile sensor | Virtual touchpad | Detection of severity of PD using a multimodal interface (Virtual touchpad) based on the UPDRS test and image analysis | Easy and reliable, doesn't require an additional hardware attached to hand | A separate camera with a stand is needed to be placed at a height to discard un-wanted objects from its active area |
| [Surangsrirat and Thanawattano 2012] | --------------- | ---------------- | Android application for Spiral analysis in PD | Archimedean and Octagon spirals used are inexpensive and easy to use and can be used for screening at home | The inexperience of working on computer devices may affect results |
| [Lin et al. 2012] | GPS | Smartphone is GPS equipped | provide personalized assistive support for elderly with demen-tia travelling outdoor | Easy to use as GPS is embedded in mobile so no need to wear other sensors which may be uncomfortable | User may sometimes forget to carry mobile with them also elderly people don't like to carry it much often |
| [Mariani et al. 2013] | TAA and gyroscope | Shoes | On-shoe wearable sensors for Gait and turning assessment of patients with PD | Practical to use in home or clinics without any discomfort for the subjects. | Person may not wear shoes most of the time |
| [Hu et al. 2013] | Implantable pulse generator, TAA | Implanted within body, Wrist | Real time and Wireless wrist-wearable wake/sleep identify-cation device on PD patients | RF module was only turned on when needed which saves power consumption | The communication distance is maximum 2m between wrist and pulse generator need to be implanted |

**Table 7: Child Healthcare Research Summary**

| Author | Sensor | Placement | Problem addressed | Advantages | Disadvantages |
|---|---|---|---|---|---|
| [Starida et al. 2003] | ---------------- | ---------------- | Healthcare monitoring at home for chronically ill children | Privacy and Security of data provided by user authentication (digital certificate), user authorization, secure transmission (S-HTTP and SSL), | High bandwidth is required for audio/video conferencing which is still not available at many places in the world |
| [Fujiwara et al. 2011] | --------------- | ---------------- | Child care training system for child minders to improve care quality | Overhead and subjective views are presented to the user for under-standing position of child free-play | Simulating actual behavior may be difficult |
| [Cann et al. 2012] | ---------------- | ---------------- | Interactive online assess - ment system for children with chronic pain | Online system is very convenient for patients and also increases the efficiency of clinician | Objective questions may not always describe the feelings patient want to convey |
| [Chuah and DiBlasio 2012] | TAA, GPS | Pants pocket, wrist | Smartphone based system to help autistic children | No extra hardware other than phone is needed to carry with autistic child | Disabled child may not carry the mobile and may not wear the device on wrist if he doesnot like it |
| [Pruette et al. 2013] | BP monitor, Webcam | ---------------- | Mobile BP telemanage-ment system for children with hypertension | Monitor BP symptoms and provide education on Hypertension in the home environment | Child may not handle equipment himself so a caretaker may be needed |

**Table 8: Women Healthcare Research Summary**

| Author | Sensor | Placement | Problem addressed | Advantages | Disadvantages |
|---|---|---|---|---|---|
| [Kanno et al. 2007] | temperature logger contains 3 temperature sensor | Abdomen | Calculates menstrual cycle, next ovulation period, using basal body temperature (BBT) and supply self-healthcare services | As system uses BBT during sleeping time so it will not interfere in normal activities | lot of images files are generated causing difficulty in processing due to which all the FOMA900 series phones also not supported |
| [Oda et al. 2007] | Temperature sensor | Abdomen | Women healthcare using basal body temperature (BBT) | User does not need to manage their data according to some specifications provided by some manufacturer of thermometer | Placement of sensors to some inaccurate position during sleep may affect the results |
| [Mazurowski et al. 2007] | ------------------ | --------------- | Knowledge based Computer assisted decision system (KB-CAD) for Breast cancer detection | The computationally expensive problem of finding mutual information between images is alleviated by reducing case-base size | Slow response time of the system when new query is presented due to computationally expensive process of calculating mutual information between images |
| [Talib et al. 2008] | ------------------- | ---------------- | Educational system for Breast cancer early detection using Tele-consultation a real time application | May provide extensive care at underserved places at low cost and also useful for emergency cases | Legal, privacy and data security concern. High bandwidth is required for Tele-consultation |
| [L. Chen et al. 2011] | Wearable BP, blood sugar, body tempera-ture, and HR | Sensor worn by pregnant woman | Maternal (Pregnant woman) and child health care intelligent system | Easily work in outdoor. Different modules are developed to monitor acute and chronic diseases. Reminds to take medication timely. | Huge computing power and storage capacity is needed to handle the flood of data and take appropriate actions |
| [Marques et al. 2000] | Sonoscope (used by preg-nant woman) | Portable foetal HR detector | Pregnancy monitoring, Remote foetal HR acquisition and analysis | Signal analysis received from sonoscope is performed in real time | Cannot be used outdoor as sonoscope is always connected to computer soundcard or telephone |

**Table 9: CVD Research Summary**

| Author | Sensor | Placement | Problem addressed | Advantages | Disadvantages |
|---|---|---|---|---|---|
| [Chen et al. 2007] | Wireless ECG sensor, 3D accelerometer | ECG sensors worn by person, accelerometer is embedded in mobile | Online ECG processing based on cellular phone for ambula-tory and continuous detection | Only abnormal ECG data is sent– reduces cost and network congestion | Processing all data in mobile phone may result into delay in emergency cases |
| [Ho and Chen 2009] | Wireless ECG sensor, 3D accelerometer | ECG sensors worn by person, accelerometer is embedded in mobile | A portable handheld cardiac health monitoring, exercise monitoring, tracking and recommendation system | System includes both exercise and cardiac health monitoring with real-time feedback which helps user | On stopping exercise user need to answer the questionnaire, which may be irritating at times |
| [Z. J. Z. Jin et al. 2009] | ECG sensor, TAA, GPS | ChestBand with ECG sensor, accelerometer is worn on chest | A personalized medical tech-nology for cardiovascular dis-ease detection & prevention | Automatically provides real time feedback detecting any abnormal heart condition | For the real time feedback user always need to wear ChestBand and need to carry mobile phone |
| [Klug et al. 2010] | ---------------- | ---------------- | Displaying Computerized ECG recordings and vital signs on windows phone 7 Smartphones (WIN7) | WIN7 with capacitive touch screen is easier to operate than older ones. Real time HR data will be sent to central monitoring station. | Viewing ECG on small smart phone display is difficult to see for elderly people. WIN7 does not support useful TCP socket. |
| [Lee et al. 2011] | 3-D Electrodes array | Cardiac-marker moni-toring system contain-ing sensors is mounte-d on patient's forearm | Design and development of mobile cardiac marker monit-oring system for prevention of myocardial infarction | Non-invasive, lightweight, small and real-time monitoring - results can be produced within 15 minutes | This system is a POCT (point-of-care-testing) device which can take rapid measurements but often results are not accurate |

**Table 10: Sleep Apnea Research Summary**

| Author | Sensor | Place-ment | Problem addressed | Advantages | Disadvantages | Subject & Result |
|---|---|---|---|---|---|---|
| [Oliver and Flores-Mangas 2007] | Blood oximeter | Wear-able during sleep | Wearable system for automatic Sleep apnea detection and monitoring with a mobile phone | light-weight, wireless, non-intrusive and allows real time processing | Need to carry gadgets like sensing module with batte-ries, DSP unit, Bluetooth module, blood oximeter. | 20 subjects wore it for full night and system identified 3 known cases with 100% accuracy. |
| [Zhang et al. 2013] | Pulse oximeter, Android smartphone | Finger-tip | A real time auto-adjust-able smart pillow system for sleep apnea detection and treatment using | Non-invasive, inexpensi-ve and portable system, First system to detect and facilitate reco-very | Special pillow is needed for detection and treat-ment of disorder. Sensor worn in the night may not be | 40 volunteers over 80 nights. The results indicate that sleep apnea duration and the numb-er of sleep apnea events reduc- |
| | | | smartphone | from sleep apnea | comfortable for some users | ed by 55% & 57% respectively. |

**Table 11: Stress Care Research Summary**

| References | Physiological Signals | Algorithms/ Techniques |
|---|---|---|
| [Sun et al. 2012] | ECG, GSR, accelerometer | J48 decision tree, Bayes network and support vector machine (SVM), K-Means clustering |
| [Setz et al. 2010] | EDA, GSR | SVM with RBF kernel (linear, quadratic and polynomial), Linear discriminant analysis (LDA), Nearest class center (NCC) |
| [Zhai and Barreto 2006] | GSR, BVP, Pupil dilation acceleration (PDA), Skin temperature (ST) | Support vector machine (SVM) |
| [Choi and Gutierrez-Osuna 2009] | HR variability (HRV) | Principal dynamic mode (PDM) |
| [de Santos et al. 2011][A. D. S. Sierra et al. 2011][A. de S. Sierra et al. 2011] | GSR, HR | Fuzzy logic |
| [Singh et al. 2011] | GSR, PPG, HR variability (HRV) | ------------------------------- |
| [Mokhayeri and Akbarzadeh-T 2011] | ECG, PPG, Pupil dilation acceleration (PDA) | Fuzzy SVM, Hough Transform |
| [Healey and Picard 2005] | ECG), EMG, GSR, and respiration | ------------------------------- |
| [Lu et al. 2012] | GSR, Microphone signals | GMM, K-means |
| [Healey 2000] | EMG, Electrodermal response or GSR, EKG or ECG, BVP, Electroencephalograph (EEG), | Fisher Analysis |
| [Lin and Hu 2005] | GSR, BVP, HR | ANOVA analysis |
| [Zhai et al. 2005] | GSR, BVP | Support vector machine (SVM) |
| [Prendinger et al. 2005] | GSR, BVP | -------------------------------------------- |
| [de Santos Sierra et al. 2010] | GSR, HR | Fisher Analysis, K-NN |
| [Katsis et al. 2006] | GSR, facial EMG, Respiration, ECG | SVM with radial basic function (RBF) kernel |
| [Liao et al. 2005] | HR, Skin temperature, GSR, Finger pressure | Dynamic Bayesian Network (DBN) |

**Table 12: Obesity Care Research Summary**

| Author | Sensor | Placement | Problem addressed | Advantages | Disadvantages |
|---|---|---|---|---|---|
| [Amft et al. 2005] | inertial sensor, wear-able motion sensors (accelerometer / gyroscope) | wrist , upper arm | Arm gesture detection related to meal intake, for automatic diet monitoring | System can indicate unhealthy eating habits like food intake with high speed or inadequate schedule | Gestures cannot indicate the calorie intake which poses a challenge to use this system for automatic diet monitoring |
| [Oliveira and Oliver 2008] | ECG, TAA | Chest | Mobile phone based system to assist runners in achieving predefined exercise goals with musical feedback and persuasion techniques | Providing real-time musical feedback and establishment of virtual competition with other runner guides and motivates user to achieve goal during their workout. | It may not be easy for user to have a look on interface providing real time feedback during workout and some user don't use phone during workout worrying about getting damaged it |
| [Consolvo et al. 2008] | 3-D accelerometer, barometer, humidity sensor, temperature sensor, visible and infrared sensor, microphone, compass | Waist | An on-body sensing, activity inference, and personal mobile system to encourage individuals to be physically active | user can manually edit the entries for which device was not either trained to infer or was trained but failed to infer | MSP's size is little bit large and is a separate device other than mobile so user may feel uncomfortable to wear it always |
| [Järvinen et al. 2008] | Light sensor | Barcode scanner uses light sensor | HyperFit: An Internet service for personal management of nutrition and exercise for promoting healthy diet and exercise | using mobile's camera as a barcode scanner makes system portable | Maintaining database of each food item is difficult and system poses a problem with those food items which don't contain barcode |
| [Sha et al. 2008] | Pulse oximeter, BP monitor, actigraph, GPS, accelerometer, sound, light, temper-ature sensor, Finger Clip Sensor | wearable sensors placed at different body parts and phone put into pants pocket | A smartphone assisted chronic illness self-management system with participatory sensing | Data is always encrypted by simple algorithms because of limited processing capability of smartphone while effectively maintaining the security and confidentiality of data | Environment sensors which are not embedded in mobile can't monitor user every time because of location constraints |
| [Gao et al. 2009] | Accelerometer, camera, GPS | Mounted at the waist or put in a pocket | Health Aware: Tackling Obesity with smart phone systems | User need to manually enter name of food item only and system will calculate calorie itself based on pre specified database | Centralized database server is used and so smartphone will not be standalone a health aware system |
| [Denning et al. 2009] | 3-D accelerometer, barometer, light sensors, humidity sensor, sound sensor, GPS | Waist | Lifestyle activity and nutrition detection using smartphone | System can automatically detect daily activities and find the calorie expenditure which will provide feedback to user to encourage him for healthy lifestyle | User need to enter data manually for the food eaten via the interface on mobile phone |
| [Khan and Siddiqi 2012] | TAA | phone placed in front pocket of pants | Promoting healthier life style using activity aware smart phones | No external sensor is used so no need to carry some extra sensor devices during performing different activities | Continuous running of activity recognition process in background may affect the battery life drastically |

**Table 13: Disabled Healthcare Research Summary**

| Author | Sensor | Placement | Communication | Problem addressed | Advantages | Disadvantages |
|---|---|---|---|---|---|---|
| [Takami et al. 1996] | Video sensor, Pressure sensor | Wear glasses having 3 LED's on it | ---------------- | Computer interface to use head movement for disabled | Allows to interact with the computer without typing on the keyboard | System requires the subject to move his head intentionally |
| [Ishimatsu et al. 1997] | Optical sensor | Wear glasses or hair-band with optical sensor on it | ---------------- | Computer interface to use head movement for disabled | Allows to control the switching on and off of the electrical devices | Patient with spinal problem may not be able to turn head |
| [Song et al. 1998] | force/torque sensor, vision sensor | Mounted on the end-effector of the robotic arm of KARES | ----------------- | Assisted living for disabled and elderly with spinal cord injuries | Simple voice commands are used to operate robotic arm | Vibration of the robotic arm may pose a problem for autonomous tasks |
| [Krasij et al. 1999] | --------------- | ---------------- | Speakers | Allows non-verbal persons to communicate | Laptop comes with audio, video hardware - not necessary to put external devices | Have to carry whole setup (not much portable) as laptop runs for 2-3 hours on battery power |
| [Roberts et al. 1999] | Infrared sensor | Sensors placed in devices to be controlled | IR technology | Home environment and entertainment remote controller for client with Spinal Bifida | Room temperature, curtains, and Stereo are controlled using a single remote controller | IR remote controls use light, so require line of sight to operate the destination device |
| [Stickney et al. 1999] | Ultrasonic sensors | 4 sensor placed in each corner of vehicle and 2 in the front | Joystick | Remotely operated toy vehicle for disabled child | Child can enjoy the toy vehicle though cannot control it himself | Another person is needed to control the vehicle remotely |
| [Carvalho et al. 1999] | Voice sensor | fitted to a standard room thermostat | Microphone | Allows voice-controlling of room temperature for Multiple Sclerosis and paraplegia patients | Allow him to control room temperature via a voice-actuated device | Command to change the temperature is in pre-specified format |
| [Kwon and Kim 1999] | Electrodes | Placed on glass frame on 5 different points in contact with skin | RF Wireless Transmitter Receiver | Wireless mouse for people with severe motor disabilities | Device is convenient to wear | May not be able to deal with subjects who spend more time for fine adjustments to cursor position |
| [Tartamella et al. 1999] | Temperature sensor | Client's wheelchair | RF Wireless Transmitter Receiver | Remote control digital thermostat and door opener for disabled | Both devices thermostat and door opener controlled through same device | Expensive to install |
| [Alecsandru et al. 1999] | RF sensors | RF sensors placed in electric appliances | RF at a frequency of about 418MHz | Remote control device for persons suffering from to control household lighting | All devices controlled by one remote control | Devices can only be switched on and off – no speed changing (e.g, fan) possible |
| [Khan and Enderle 2000] | Temperature sensor inside thermostat | Fitted to wheelchair but can be detached & operated from anywhere inside apartment | RF Wireless Transmitter Receiver | System for patients suffering from Reflex Sympathetic dystrophy syndrome | Alternate power backup, Multiple security codes to prevent interference with other devices | Expensive to install |
| [Hela et al. 2001] | GPS | Head-mounted display for visual tracking (disabled), integrated headset for speech I/O (blind), GPS in the backpack | 802.11b WLAN | Wireless pedestrian (outdoor) navigation system for visually impaired and disabled | Blind user can add intelligence, as perceived, to the central server hosting the spatial database | Less signal strength near tall buildings and under tree canopies so can't be used inside buildings |
| [Ran et al. 2004] | Ultrasound tag, GPS | Head-mounted display for visual tracking (disabled), integrated headset for speech I/O (blind), Ultrasound tag attached to both shoulders | 802.11b WLAN | Wireless indoor and outdoor navigation system for visually impaired and disabled | System provides both indoor and outdoor navigation in single device | Range of wireless network determines the working range of system. No automatic switching between indoor and outdoor modes |
| [Ding et al. 2005] | Electrodes | 2 electrodes to each of the orbits of the eye & 1 to the bridge of nose | EOG (Electro-oculography) signals | Allows disabled persons to communicate using eye as input device | Non-invasive, low cost and easy to use, No direct attachment to eyes | Not the most accurate technology to detect eye movement |
| [Liu et al. 2005] | Microphone | microphone connected to computer | Ethernet or wireless connection | Handle computers /robots with voice command | Better than remote control for the users feeling pain in touching objects | Pronunciation of user may affect the functioning |
| [Lim and Singh 2005] | Wearable eyes sensor | Spectacles | Bluetooth | Communication using eye blink detection system for ALS patients | ALS (diseases with no muscle movement) patients can communicate through eye blinking | Patient always need computer to ommunicate, Normal blinking of eye can pose a problem |
| [Tanaka et al. 2008] | CMOS Image Sensor, LED | 2 LED's attached to head in one device, 2 LED's attached to mouth in second device | IR signals | Vision and slight body movement based pointing device for people with ALS and cerebral infarction | Intentional movement of the body can be extracted with acceptable speed and accuracy | 11-lettered-word was wri-tten on average in 53 and 69 secs by 5 subjects with 2 devices, respectively |
| [Lv et al. 2008] | 16 Ultrasonic sensor | Robot | WLAN | Handle computers or robots with voice command | 16 ultrasonic sensors are used to detect obstacle in all directions | In noisy environment efficiency of system drops |
| [Su et al. 2008] | Video sensor | Video camera placed in front of computer | ------------------- | Allows people with severe disabilities to access com-puter and to communicate using eye blinks | No sensor needed to wear on eye | If eye template updating is not timely w.r.to envir-onmental changes, it may become outdated and may locate wrong eye regions |
| [Lee and Lee 2009] | 4 Ultrasound, HR monitor, 4 vibration sensor, | HR monitor on chest/ wrist, ultrasound and RF transmitters on head, ultrasound sate-llite on ceiling, Vibra-tion sensor on belt | RF, UHF, Ultrasound, WSN | Indoor positioning system for moving objects and playing sports for blind or visually impaired | This system makes it easy to play indoor games and also monitors the real time HR | Visually impaired person can play only indoor games with lot of setup and wearable devices attached to the body |
| [Krejcar 2011] | CCD sensor | CCD (Charge-Coupled Device) is inside camera | ------------------- | Virtual keyboard for handicapped controlled by head motion & eyelids | Cheaper than existing systems, no extra hard-ware needed | High error rate in poor lighting conditions while writing a text |
| [L.J.G. et al. 2011] | TAA as head tilt sensor | Head mounting system integrate accelerometer | IR LED's | Computer interface for disabled using voluntary eye movement | TAA compensate head tilt positions efficiently and accurately | Not all the head tilt positions are considered |
| [El Alamy et al. 2012] | RFID tags | RFID tag integrated in PDA, RFID reader in bus station | Bluetooth, Wi-Fi | Bus Identification System for Visually Impaired Person | Only bus station has RFID reader which then send signals to all buses | Braille keyboard is too costly (approx.. $1000) |
| [Ganz et al. 2012] | RFID tags | RFID tags deployed in each room | Bluetooth, Wi-Fi | Indoor navigation for visually impaired | The system is scalable to any size building | User need to carry smart-phone and PERCEPT Glove which gets discharged within 2 hours |
| [Subbu and Gnanaraj 2012] | Vibration motor | 6 Vibration motors (3X2) form refreshable Braille display | Bluetooth, Braille code | Electronic device to enable visually impaired to read messages from modern gadgets | Braille can be translated to any language of the world | A separate device is need-ed for reading messages, that is less compact and so not easily portable |
| [Chuah and DiBlasio 2012] | TAA, GPS | Pants pocket, wrist | Bluetooth, Wi-Fi | Smartphone based system to monitor autistic children | No extra hardware other than phone is needed to carry with autistic child | Disabled child may not want to carry the phone or to wear the device on wrist if he does not like it |
| [Yu et al. 2013] | 6 ultrasonic sensors | Sensors placed in cart | Tactile simulator | Safe walking guide system for visually impaired person with recognition of obstacles | Gives feedback to the user through vibro-tactile stim-ulation to the palm which no earlier system provides | Experiment conditions are limited to indoor envi-ronment with plain terra-in and fixed obstacles |

**Table 14: Commercial Pervasive Healthcare Products**

| App Name | Sensor | Hardware / Software Support | Problem addressed | Extra Features | Limitations |
|---|---|---|---|---|---|
| Nike+ Running | GPS, ace-lerometer | iOS | Fitness | Record your distance, pace, and time. In-run audio feedback at every mile. Tag Facebook friends share a route map with. | Continued use of GPS running in the background could dramatically decrease battery life. |
| Instant Heart Rate | Phone's camera (place finger on camera) | iOS | HR monitoring | Real-time PPG graph. Continuous or Auto-Stop mode. One week data storage and tags sharing to Twitter and Facebook | Fingertip should completely cover the camera lens for accurate results. Pressing too hard reduces blood flow so press fingertip gently. |
| Cardiograph | Phone's Camera | iOS | HR Pulse Measurement | Calculate heart's rhythm with pictures of your fingertip. Can save your results for future reference, for sharing or safe keeping. | In device without built-in camera flash, measurements should be taken in a well-lit environment. |
| Run-Keeper | GPS | iOS | GPS Track Running Walking Cycling | Track your fitness activities. Share with friends on Facebook, twitter on RunKeeper. | Continuously running GPS in the back-ground significantly drains battery |
| Sleep Time | accelerometer | iOS | Analyzes sleep to wake up user at the perfect moment of lightest sleep phase | Alarm Clock and Sleep Cycle Analysis with Soundscapes | The app uses a 3rd party service Flurry Analytics, which uses the device MAC address to track the app's usage pattern |
| Sleep Cycle | accelerometer | iOS | Bio alarm clock that wakes you in lightest sleep phase | It finds the optimal time to wake you up during a 30 minute window that ends at your set alarm time. | Phone's placement should be checked before sleeping. |
| Stress Check | Phone's camera | iOS | Stress check to manage your emotional and physical stress level. | By measuring HR variability(HRV) using iPhone's built-in camera, level of stress can be estimated in just 2 minutes anywhere, anytime. | Phone's placement should be checked before sleeping. |
| Runtastic Heart Rate PRO & Pulse Monitor | iPhone camera | iOS | HR measurement at anytime, anywhere. | User can see the development of HR in a graph. Automatic reminder to measure your HR. Share results on Facebook, Twitter. | Be calm, Place and lightly hold index finger gently against the back camera lens for accurate result. Do not measure with cold fingers. |
| Runtastic PRO | GPS | iOS | Track workout time, distance, speed, eleva-tion, calories, and more | Audio feedback for each mile or km completed. Auto Pause when stop running. Geo Tagging | Continued use of GPS running in the background can dramatically decrease battery life. |
| Run with Map My Run | GPS | iOS | Running, Jog, Walk, Workout Tracking and Calorie Counter | Calculates route, mileage, minutes per mile and calories burnt. Great for tracking both indoor and outdoor activities | Turns off Wi-Fi to improve battery life and GPS accuracy. Continued use of GPS can drain battery completely. |
| Sports Tracker | GPS | iOS | Track and analyze your performance, get fit and stay healthier | Keeps track of calories burnt, average speed, and altitude. Dedicated Indoor mode with HR features. Share workout data and photos on Sports Tracker, Facebook and Twitter | Continued use of GPS can dramati-cally decrease battery life. iPod Touch devices do not have internal GPS, so the location accuracy is very limited. |
| Strava Cycling | GPS, ANT+, Bluetooth low energy (BLE) | iOS | Biking and Riding Route Tracker | See distance, speed, elevation gained, and calories burnt. Record and view the map of ride. Collect HR, power and cadence data from sensors (supports ANT+ and Bluetooth LE) | Continued use of GPS running in the background can dramatically decrease battery life. |
| iBGStar Blood Glucose monitoring system | ---------------- | iOS | Manage Blood Glucose By drawing a bit (0.5 microliter) of blood. | The app allows tracking glucose, insulin, and carbs and gives various ways to view that data—from charts and graphs to results comparisons by time-of-day. | A separate 30-pin adapter needed to carry with iPhone. Inaccurate result may be due to not properly washing or drying hands, dehydration |
| Endomondo Sports Tracker PRO | GPS, Pedometer, ANT+ sensors | Android, Blackberry, iOS | Track outdoor sports activities | Audio feedback, see tracking route on map, Share on social network, Track your HR | Continuous GPS may decrease battery life. |
| Sportypal | GPS | Android, Blackberry, iOS, Java, Symbian | Sports training and tracking app | Auto pause, Auto start, Share on Facebook and Twitter | Continuous GPS may decrease battery life. |
| Cardiio | Camera | iOS | Touch free HR monitoring | No need to touch the camera to monitor HR. Analyzes HR data to provide a fitness level rating and also estimates potential life expectancy. | ---------------------- |